\documentclass[aps,superscriptaddress,eqsecnum,nofootinbib,showpacs,preprintnumbers]{revtex4}
\usepackage[utf8]{inputenc}
\usepackage{hyperref}
\hypersetup{
    colorlinks=true,
    linkcolor=blue,
    filecolor=magenta,
    urlcolor=cyan,
}
\usepackage{amsmath}
\usepackage{amsfonts}
\usepackage{latexsym}
\usepackage{xcolor}
\usepackage{graphicx}
\usepackage{float}
\usepackage{xcolor}
\usepackage{amssymb}
\usepackage{graphicx, rotating}
\usepackage{epsfig}
\usepackage{latexsym}
\usepackage{graphicx}
\usepackage{color}
\usepackage{amsmath,bm,amssymb}
\hypersetup{colorlinks=true, citecolor=bluscuro, linkcolor=black, urlcolor=bluscuro}
\definecolor{rossos}{cmyk}{0,1,1,0.55}
\definecolor{bluscuro}{rgb}{0.15, 0.2, .85}
\definecolor{bluchiaro}{cmyk}{1,.3,0.,0.1}
\newcommand{\be}{\begin{equation}}
\newcommand{\ee}{\end{equation}}

\usepackage{amsmath}
\numberwithin{equation}{section}

\def\ba{\begin{eqnarray}}
\def\ea{\end{eqnarray}}

\def\bq{\begin{quote}}
\def\eq{\end{quote}}

\newcommand{\pp}{\prime \prime}

\newcommand{\f}{\frac}

 at 10truept

\newcommand{\beq}{\begin{equation}}
\newcommand{\eeq}{\end{equation}}
\newcommand{\beqa}{\begin{eqnarray}}
\newcommand{\eeqa}{\end{eqnarray}}
\newcommand{\bea}{\begin{eqnarray}}
\newcommand{\eea}{\end{eqnarray}}
\newcommand{\p}{\partial}

\def\ltap{\ \raise.3ex\hbox{$<$\kern-.75em\lower1ex\hbox{$\sim$}}\ }
\def\gtap{\ \raise.3ex\hbox{$>$\kern-.75em\lower1ex\hbox{$\sim$}}\ }
\def\gl{\ \raise.5ex\hbox{$>$}\kern-.8em\lower.5ex\hbox{$<$}\ }
\def\roughly#1{\raise.3ex\hbox{$#1$\kern-.75em\lower1ex\hbox{$\sim$}}}

\allowdisplaybreaks

\def\pp{{\prime\prime}}
\def\pr{{\prime}}

\begin{document}

\title{Topological Black Holes with curvature induced scalarization in the extended scalar-tensor theories}
	
	\author{Stella Kiorpelidi}
	\email{<stellakiorp@windowslive.com>}
	\affiliation{Physics Department, National Technical University of Athens, 15780 Zografou Campus, Athens, Greece}
	
	\author{George Koutsoumbas}
	\email{kutsubas@central.ntua.gr}
	\affiliation{Physics Department, National Technical University of Athens, 15780 Zografou Campus, Athens, Greece}

\author{Andri Machattou}
	\email{<andrimachattou@hotmail.com>}
	\affiliation{Physics Department, National Technical University of Athens, 15780 Zografou Campus, Athens, Greece}
	
	\author{Eleftherios Papantonopoulos}
	\email{lpapa@central.ntua.gr} \affiliation{Physics Department, National Technical University of Athens, 15780 Zografou Campus, Athens, Greece}
	
\begin{abstract}
We study the perturbative behaviour of topological black holes in the presence of a cosmological constant and a scalar field coupled to the Gauss-Bonnet term. We calculate both analytically and numerically the quasi-normal modes of scalar perturbations in the extended scalar-tensor-Gauss-Bonnet gravity. In the case of small black holes we find  a phase transition of the topological black hole to a hairy configuration.
\end{abstract}

\maketitle
\tableofcontents

\section{Introduction}

The recent experimental results on gravitational waves \cite{GW1,GW2,GW3} and more recently the observation of a  shadow of the $M87$ black hole \cite{EHT}, demonstrated that Einstein's General Relativity (GR) is a very successful viable theory. However on cosmological grounds, to explain the recent observational results on dark matter and on dark energy a generalization of GR is required, in a attempt to have a viable theory of Gravity on short and large distances \cite{Joyce:2014kja,Nojiri:2006ri,Clifton:2011jh,MG3}. These modified gravity theories can give us important information on the structure and properties of the compact objects predicted by these theories and also  the observational signatures, which they can introduce.

Some of the  simplest and viable modifications of GR are the scalar-tensor theories \cite{Fujii}. When the scalar field coupled to gravity backreacts to the background metric, hairy black hole solutions would be generated. A hairy black hole solution in an asymptotically flat spacetime was found in \cite{BBMB} but it was shown that it was unstable because the scalar field was divergent on the event horizon \cite{bronnikov}. However, it was soon realized that introducing  a scale through the presence of a   cosmological constant, making the space-time asymptotically AdS/dS, such an irregular behaviour of the scalar field  on the horizon was avoided. Then hairy black hole solutions were found having  a regular scalar field behaviour and all the possible divergence were hidden behind the horizon \cite{Martinez:1996gn,Banados:1992wn,Martinez:2004nb,Zloshchastiev:2004ny,Martinez:2002ru,
Dotti:2007cp,Torii:2001pg,Winstanley:2002jt,Martinez:2006an,Kolyvaris:2009pc,Charmousis:2014zaa}.

If the cosmological constant is positive and the scalar field is minimally coupled or non-minimally coupled  with a self-interaction potential,
black hole solutions were found \cite{Zloshchastiev:2004ny,Torii:1998ir,Martinez:2002ru} but it was shown to be unstable \cite{Harper:2003wt,Dotti:2007cp}. If the cosmological constant is negative numerical solutions were found  \cite{Torii:2001pg,Winstanley:2002jt} and also a stable exact black hole solution was discussed in \cite{Martinez:2004nb} in which the space-time is asymptotically AdS with hyperbolic geometry, known as MTZ black hole. Later this solution was generalized to include charge
\cite{Martinez:2005di} while a generalization to
non-conformal solutions was discussed in \cite{Kolyvaris:2009pc}.

No-hair theorems  can also be evaded by considering black holes interacting with  matter fields \cite{Stefanov_2008}-\cite{Myung2018}.
In such cases black holes can support a non-trivial scalar field in their exterior region. Modified gravity theories were proposed in which matter is coupled to the Einstein tensor. These theories belong to  general scalar-tensor Horndeski theories \cite{Horndeski}. Then various hairy black holes were found in which scalar fields are coupled to curvature \cite{Kolyvaris:2011fk, Rinaldi:2012vy,
Kolyvaris:2013zfa, Babichev:2013cya, Cisterna:2014nua,  Charmousis:2014zaa, Koutsoumbas:2015ekk, Anabalon:2013, Cisterna:2015, Cisterna:2016}.

Hairy black hole solutions can also be obtained without the presence of matter sources if the scalar field is directly coupled to second order algebraic curvature invariants. In this case the scalar hair is maintained by the interaction with the spacetime curvature. Exploring the strong field regime
of gravity with the aim to detect  gravitational waves and black hole shadows the effects of higher-order curvature terms become  significant. However, including such terms brings in the well-known ghost problem \cite{stelle}. One high curvature correction is the Gauss-Bonnet (GB) term  which is ghost-free but it becomes a topological term in four-dimensional spacetime and has
no dynamics. To evade this problem, one has to couple this term to a scalar field in four dimensions \cite{stringT}. These gravity theories are known as extended scalar-tensor-Gauss-Bonnet (ESTGB) theories and  were studied extensively in the literature \cite{Mignemi_1993}-\cite{Kleihaus_2016a}.

Recently there is a lot of activity studying  the ESTGB gravity theories in an attempt to evade  the no-hair theorems  and obtained hairy black hole solutions.  In particular, for certain classes of the coupling function it was shown that we have spontaneous scalarization of black holes \cite{Doneva_2018a}-\cite{Antoniou_2018a}. It was found  that below a certain critical mass the Schwarzschild black hole becomes unstable in regions
of strong curvature, and then when the scalar field backreacts to the metric,   new branches of scalarized black holes develop at certain masses as solutions in the ESTBG theory \cite{Doneva_2018a,Silva_2018,Myung_2018b}.  An extension of these results is to consider the case of nonzero black hole charge. Examining  the entropy of the black holes with nontrivial scalar field it turned out that the solution with the scalar field is   thermodynamically favorable over the Reissner-Nordstr\"{o}m one \cite{Doneva:2018rou}.

The spontaneous scalarization procedure has various applications. The scalarization due to a coupling of a scalar field to Ricci scalar was studied in \cite{Herdeiro:2019yjy} and scalarized black hole solutions and compact objects in asymptotical flat spacetime in the ESTGB gravity theories were obtained in \cite{Silva:2017uqg}-\cite{Hunter:2020wkd} and also in AdS/dS spacetimes \cite{Bakopoulos:2018nui,Brihaye:2019gla,Bakopoulos:2019tvc,Bakopoulos:2020dfg,Lin:2020asf}. The connections of asymptotically AdS  black holes  scalarization  with  holographic phase transitions in the dual boundary theory was studied in \cite{Brihaye:2019dck,Guo:2020sdu}. Recently the spontaneous scalarization in f(R) gravity theories  was discussed in \cite{Tang:2020sjs}.

The black hole spontaneous scalarization in ESTGB gravity theories with a probe scalar field in a black hole background with different curvature topologies has been studied in \cite{Guo:2020zqm}. It was found that the scalar field near AdS black hole with positive curvature could be much easier to scalarize the black hole comparing with negative and zero curvature cases. In particular, when the  curvature is negative, the scalar field is the most difficult to be bounded near the horizon. It was observed that scalarizations in hyperbolic AdS topological black hole (TBH) backgrounds depend  on the interplay of two factors, the  coupling  strength  between  the  scalar  field  and  the  GB  term  and  the cosmological constant.

As we already mentioned, the MTZ black hole \cite{Martinez:2004nb} is an exact black hole solution in four dimensions with a minimally coupled self-interacting scalar field, in an asymptotically  AdS space-time in which the  event horizon is a surface of negative constant curvature enclosing the curvature singularity. It was shown that there is a second order phase transition at a critical temperature below which a black hole in vacuum undergoes a spontaneous dressing up with a non trivial scalar field. In a series of papers \cite{Koutsoumbas:2006xj,Koutsoumbas:2008pw,Koutsoumbas:2008yq} this scalarization procedure was studied for topological black holes. Calculating analytically and numerically the quasinormal modes (QNMs) of tensor, electromagnetic and scalar perturbations, it was found that there is a critical value of the  horizon radius  below which  the topological black hole is scalarized to the  MTZ black hole with scalar hair. The thermodynamics of this transition was also studied.

Motivated by the above studies we will study the scalarization of a topological black hole in the presence of the coupling of the scalar field to the GB term in  the ESTGB gravity theories. In particular we will consider a gravity theory with the presence of a cosmological constant in which there is matter parametrized by a massive scalar field minimally coupled to gravity and also coupled to the GB term. The coupling of the scalar field to the GB term is denoted by the parameter $\lambda$. At first place the scalar field does not back-react to the metric. We fix the background metric to be a TBH leaving in a hyperbolic space-time expressed by a parameter $\xi,$ which is analogous of the orbital quantum number in the three-dimensional space.

The goal of this work  is to study the behaviour of matter in this physical set up. For fixed cosmological constant we have two competing effects. The first one is as $\lambda$ is increasing we expect the matter to interact more strongly with gravity while as $\xi$ is getting larger the kinetic effects tend to dominate.  We calculate both analytically and numerically the
QNMs of scalar perturbations of topological-AdS black holes in the presence of matter coupled to the GB term. For each $\lambda$ we found  a critical value of $\xi,$  below which there is instability. As $\lambda$  is increasing the imaginary part of some QNMs are getting positive indicating an instability. When the coupling constant $\lambda$  is getting very large, all of the QNMs develop a positive imaginary part. This behaviour provides evidence of a phase transition to a scalarized TBH. We also noted that the absolute values of QNMs are increasing as the parameter $\xi$ is also increasing.

The work is organized as follows. In Section \ref{sec2} we present the theory of the coupling of a scalar field to the GB term in the background of a TBH and we discuss the tachyonic instabilities of this theory. In Section \ref{sec3} we carried out an analytical calculation of QNMs. In Section \ref{sec4} we consider scalar perturbations in the extended scalar-tensor GB theory in which the background metric is the TBH and finally in Section \ref{sec5} are our conclusions.

\section{Topological black holes, the Einstein-Scalar-Gauss-Bonnet theory and Tachyonic instabilities  }
\label{sec2}

In this Section we will first discuss the TBHs as the background of the scalar-GB gravity theories and then we will discuss the possible tachyonic instabilities of these theories.

We consider the bulk action
\be I=\frac{1}{16\pi G} \int
d^{4}x\sqrt{-g}\left[ R+\frac{6}{l^{2}}\right]~,\ee
in
asymptotically AdS spacetime  where $G$ is the Newton's constant and
$l$ is the AdS radius. The presence of
 a negative cosmological constant $(\Lambda=-\frac{3}{l^{2}})$
  allows the existence of black holes with a
topology $\mathbb{R} \times \Sigma$, where $\Sigma $ is a
two-dimensional manifold of constant negative curvature. These
black holes are known as topological black holes. The simplest
solution of this kind reads
\begin{equation}
ds^2 =
-g(r) dt^2 + \frac{dr^2}{g(r)} + r^2 d\sigma^2  \ , \ \ g(r) = r^2
- 1 - \frac{2\mu}{r}~,  \label{Top-BH-Einstein}
\end{equation}
where we employed units in which the AdS radius is $l=1$ and
$d\sigma$ is the line element of $\Sigma $. The latter is locally
isomorphic to the hyperbolic manifold $H^2$ and of the form
\be \Sigma =H^{2}/\Gamma \quad\quad , \quad \Gamma \subset
O(2,1)\;, \ee where $\Gamma$ is a freely acting discrete
subgroup (i.e., without fixed points) of isometries.

The geometry of the TBHs  as well their basic properties
have been studied extensively in the
literature~\cite{Lemos}-\cite{Birmingham}.  It has been  shown in
\cite{Gibbons:2002pq} that the massless configurations where
$\Sigma $ has negative constant curvature are stable under
gravitational perturbations. The stability also of the
TBHs was discussed in \cite{Birmingham:2007yv} and  QNMs in topological black holes were calculated in
\cite{Wang:2001tk}-\cite{Birmingham:2006zx}.

The Einstein-Scalar-Gauss-Bonnet theory is described by the following action functional
\begin{equation}
    S=\frac{1}{16 \pi G_N} \int \, d^4x \, \sqrt{-g} \left[ R-2 \nabla_\mu \phi \nabla^\mu \phi-m^2 \phi^2 +\lambda^2 f(\phi) \mathcal{R}^2_{GB}-2\Lambda \right]~.\label{action}
\end{equation}
This modified gravitational theory consists of  a real scalar field  minimally coupled to Einstein’s gravity and non-minimally coupled to  quadratic gravitational GB term $\mathcal{R}^2_{GB}$ through a real function $f(\phi)$. A cosmological constant $\Lambda$ is also present, that may take either a positive or a negative value. We are interesting in hyperbolic TBHs with negative curvature constant. So the metric ansatz  reads as
\begin{equation}
    ds^2=-e^{A(r)}dt^2+e^{B(r)}dr^2+r^2\left(d\theta^2+\sinh^2{\theta}d\varphi^2 \right)~.  \label{hyperbolicmetricansatz}
\end{equation}
Using natural units such that $G_N=c=1$ the gravitational field equations have the covariant form
\begin{equation}
    G_{\mu\nu} =\tilde{T}_{\mu\nu}~, \label{EE}
\end{equation}
\begin{equation}
    G_{\mu\nu} = T_{\mu\nu}^{(\phi)}+T_{\mu\nu}^{(GB)}-\Lambda g_{\mu\nu}~,
\end{equation}
\begin{equation}
    G_{\mu\nu}+\Lambda g_{\mu\nu} = T_{\mu\nu}^{(\phi)}+T_{\mu\nu}^{(GB)}~, \label{fieldequations}
\end{equation}
where $G_{\mu\nu}$ is the Einstein tensor
\begin{equation}
    G_{\mu\nu}=R_{\mu\nu}-\frac{1}{2}g_{\mu\nu} R~,
\end{equation}
and $T_{\mu\nu}^{(\phi)}$ is the energy-momentum tensor which receives
contribution only from the kinetic term and the mass term of the scalar field and $T_{\mu\nu}^{(GB)}$ is the energy-momentum tensor which receives
contribution only from the interaction of the scalar field with the Gauss-Bonnet term
\begin{align}
    T_{\mu\nu}^{(\phi)}=&- \frac{1}{2} g_{\mu\nu} m^2 \phi^2+2 \nabla_\mu \phi \nabla_\nu \phi-g_{\mu\nu} \nabla_\kappa \phi \nabla^\kappa \phi~, \\
    T_{\mu\nu}^{(GB)}=& -R(\nabla_\mu \Psi_\nu+\nabla_\nu\Psi_\mu)-4\nabla^\alpha\Psi_\alpha G_{\mu\nu} \nonumber\\
    &+4R_{\mu\alpha}\nabla^\alpha\Psi_\nu +4R_{\nu\alpha}\nabla^\alpha\Psi_\mu - 4 g_{\mu\nu} R^{\alpha\beta} \nabla_\alpha \Psi_\beta~, \nonumber \\
    &+ 4 R^\beta{}_{\mu\alpha\nu} \nabla^\alpha \Psi_\beta~,  \label{TGB}
\end{align}
with
\begin{equation}
    \Psi_\mu =\lambda^2 \dot{f}(\phi)\nabla_\mu \phi~. \label{Psi}
\end{equation}
The equation of motion of the scalar field is
\begin{equation}
    \nabla_\mu \nabla^\mu \phi -\frac{1}{2}m^2 \phi+\frac{1}{4}\lambda^2 \dot{f}(\phi)\mathcal{R}^2_{GB}=0~, \label{eomscalar}
\end{equation}
where the dot denotes differentiation with respect to the scalar field. A condition for the coupling function $f(\phi)$ arises from the Eq.~(\ref{eomscalar}), namely $\dot{f}(0)=0$. This condition ensures that the trivial scalar field $(\phi=0)$ satisfies the equation of motion.
In the case of a trivial scalar field the metric functions of the background TBH are given by
\begin{equation}
   e^{A(r)}\equiv g(r) =-1-\frac{M}{r}-\frac{\Lambda}{3}r^2~,
\end{equation}
\begin{equation}
    e^{B(r)}=\frac{1}{-1-\frac{M}{r}-\frac{\Lambda}{3}r^2 }=\f{1}{g(r)}~.
\end{equation}
The equation of motion of the scalar field (\ref{eomscalar}) can be written as
\begin{equation}
    \left( \Box-\mu^2_{eff}\right) \phi=0~, \label{scal22}
\end{equation}
where
\begin{equation}
    \mu^2_{eff}=\frac{1}{2} m^2-\frac{1}{4}\lambda^2 \dot{f}(\phi) \mathcal{R}^2_{GB}~. \label{eff22}
    \end{equation}

The sign of this effective mass is connected with the stability or instability of the underlining theory. To clarify this issue consider the Lagrangian density for a free relativistic scalar field $\phi$ in a Minkowski spacetime
\begin{equation}
    \mathcal{L}=-\frac{\epsilon}{2}\partial_\mu \phi \partial^\mu \phi - \frac{\varepsilon}{2}m^2\phi^2~.
\end{equation}
In the $\epsilon=\varepsilon = +1$  case the Hamiltonian is positive semi-definite and therefore bounded from below, while in the $\epsilon=\varepsilon = -1$ case the Hamiltonian is negative semi-definite and therefore bounded from above. In the case $\epsilon = -\varepsilon$, the Hamiltonian is indefinite and so it is not bounded either from below or from above. The field $\phi$ is called a ghost field if $\epsilon=\varepsilon = -1$ (for a review on ghost fields see \cite{Sbisa:2014pzo}), while it is called a tachyon field if $\epsilon = +1$ and $\varepsilon = -1$, finally, it is called a tachyonic ghost if $\epsilon = -1$ and $\varepsilon = +1$. A Hamiltonian which is unbounded from below is usually associated with instabilities of the system. If $\epsilon=-\varepsilon$, a small perturbation can grow exponentially, signalling an instability.

If the effective mass (\ref{eff22})  is negative $ \mu^2_{eff}<0 $ there is a tachyonic instability triggered by a negative effective mass squared of the scalar field \cite{Brihaye:2019gla}.

\section{Analytical calculation of QNMs}
\label{sec3}

We consider a function $f(\phi)$ coupled to the GB term, for which
%\be f(\phi)=\f{\phi^2-2 \phi^4}{2 \phi^4+1}~.\ee
 \be \left. \f{d f}{d \phi}\right|_{\phi=0} = 0~,\ \ \ \left. \f{d^2 f}{d \phi^2}\right|_{\phi=0}=2>0~. \ee
We suppose that $\phi$ is restricted in the well surrounding $\phi=0,$ so that the expression $\lambda^2 f(\phi) {\cal R}^2_{GB(0)} $ reduces in this limit to the approximate form
 $ \lambda^2 {\cal R}^2_{GB(0)} \phi^2~.$
We consider the line element \be ds^2 = -g(r) dt^2+\f{1}{g(r)}dr^2 +r^2 (d\theta^2+r^2\sinh^2\theta d\phi^2)~,\ \ g(r)=-1-\f{M}{r} -\f{\Lambda}{3} r^2 = -1-\f{M}{r}+\f{r^2}{L^2}~,\ee for which the GB invariant reads  \be {\cal R}^2_{GB(0)} = \f{24}{L^4}+\f{12 M^2}{r^6}~.\ee
The starting point of our approach will be an equation describing the scalar perturbations, which derives from the Klein-Gordon equation after substituting for the scalar field the form
 \be \phi(t,\ r,\ \theta,\ \phi) =\Psi(r) e^{-i\omega t} {\cal Y}_{\xi m}(\theta,\phi)~,\ee where ${\cal Y}_{\xi m}(\theta,\phi)$ are the counterparts of the spherical harmonics and $\xi$ is a parameter analogous to the orbital quantum number. The equation reads \be g(r)\f{d}{d r} \left(g(r) \f{d \Psi}{d r}\right)+[\omega^2-{\cal V}(r)] \Psi=0~,\ \ {\cal V}(r) \equiv  g(r)\left[\f{g^\pr(r)}{r}-\f{\lambda^2}{4} {\cal R}^2_{GB(0)}+\f{\xi^2+\f{1}{4}}{r^2}\right]~.\label{eq1}\ee
  Note that the parameter $\xi$ indicates the hyperbolic geometry. A large value of $\xi$ shows  the departure from the spherical topology.

  Substituting $g(r)$ and ${\cal R}^2_{GB(0)}$ by their respective values, equation (\ref{eq1}) takes the form
\be  g(r)\f{d}{d r} \left(g(r) \f{d \Psi}{d r}\right) +\left[\omega^2 - g(r)\left(\f{2}{L^2} + \f{M}{r^3} - \f{6 \lambda^2}{L^4} - \f{3 \lambda^2 M^2}{r^6} +\f{\xi^2+\f{1}{4}}{r^2}\right)\right]\Psi=0~.\label{eq2}\ee
We now introduce the new variable
\be u\equiv \left(\f{r_+}{r}\right)^2\Leftrightarrow r=\f{r_+}{u^{1/2}}~,\ee
so that
\be \f{d}{dr}= -\f{2 u^{3/2}}{r_+}\f{d}{du}, \ \f{d^2}{dr^2}= \f{4}{r_+^2} u^{3/2} \f{d}{du}\left(u^{3/2}\f{d}{du}\right)~,\ee
and equation (\ref{eq2}) becomes
\be  \f{4}{r_+^2} g u^{3/2}\f{d}{d u}\left(g u^{3/2} \f{d}{du}\right)\Psi+\left[\omega^2 - g\left( 2 + \f{M u^{3/2}}{r_+^3} - 6 \lambda^2 - \f{3 \lambda^2 M^2 u^3}{r_+^6} +\f{\xi^2+\f{1}{4}}{r_+^2}u \right)\right]\Psi=0~.\ee
Setting $L=1$ and using the notations
  \be \hat{g}(u)\equiv \f{g(r)}{r_+^2} =\f{1}{u}-\f{1}{r_+^2}-\f{M}{r_+^3} u^{1/2}~,\ee
\be \hat{{\cal V}}(u)\equiv \f{{\cal V}(r)}{r_+^2} = \hat{g}(u)\left[ 2 + \f{M u^{3/2}}{r_+^3} - 6 \lambda^2 - \f{3 \lambda^2 M^2 u^3}{r_+^6} +\f{\xi^2+\f{1}{4}}{r_+^2}u\right],\ee
the equation takes the form
\be -4 u^{3/2} \hat{g}(u) (u^{3/2} \hat{g}(u) \Psi^\pr)^\pr +\hat{{\cal V}}\Psi=\hat{\omega}^2\Psi\Leftrightarrow H\Psi = \hat{\omega}^2\Psi,\ \ \hat{\omega}\equiv\f{\omega}{r_+}~.\ee
It is possible to proceed with an analytical approach, in two cases, namely when the black hole is small with $r_+$ is around $1,$ or when the black hole is large.

\subsection{Small black hole: the horizon is approximately one}

We now restrict our attention to the critical case,
where $$r_+=1\Leftrightarrow M=0~.$$
In this case $$\hat{g}(u)= \f{1-u}{u}~,$$
and the equation reduces to
\be -4 u^{1/2} \hat{f}(u) (u^{1/2} (1-u) \Psi^\pr)^\pr +\hat{{\cal V}}(u)\Psi=\f{\hat{\omega}^2}{1-u}\Psi\Leftrightarrow H\Psi = \f{\hat{\omega}^2}{1-u}\Psi~,\ee
with
  $$\hat{{\cal V(u)}}=\f{2 -6 \lambda^2}{u}+\left(\xi^2+\f{1}{4}\right)~,$$ and then the Klein-Gordon equation becomes
  \be -4 u^{1/2} (u^{1/2} (1-u) \Psi^\pr)^\pr +\left(\f{2 - 6 \lambda^2}{u}+\xi^2+\f{1}{4}\right)\Psi=\f{\hat{\omega}^2}{1-u}\Psi~. \label{eqhor}\ee
We introduce the parameter $a\equiv 2 -6 \lambda^2,$ so that the potential takes on the simple form
 $$\hat{V}=\f{a}{u}+\xi^2+\f{1}{4}~,$$
 and the equation to be solved reads
\be 4 (1-u) u \Psi^\pp + 2 (1-u) \Psi^\pr -4 u \Psi^\pr +\f{\hat{\omega}^2}{1-u}\Psi -\left(\f{a}{u}+\xi^2+\f{1}{4}\right)\Psi=0~.\ee
One may check that the (finite) approximate solution in the limit $u\to 0$ is proportional to $u^{\f{1+\sqrt{1+4 a}}{4}},$ while in the limit $u\to 1$ it is proportional to: $(1-u)^{\pm\f{i \omega}{2}}.$ We choose $(1-u)^{-\f{i \omega}{2}},$ in which case a negative  imaginary part of $\omega$ corresponds to a stable system. It is convenient to make the transformation
 \be \Psi(u)=u^{\f{1+\sqrt{1+4 a}}{4}} (1-u)^{-\f{i \omega}{2}} X(u)\label{transf}~.\ee
 The function $ X(u)$ interpolates between the two limiting values of $u\to 1$ and $u\to 0$. Then  the differential equation becomes
\be 16 u (1-u)X^\pp(u)+ 8 [ -2 -\sqrt{1+4 a} +u (4 +\sqrt{1+4 a} -2 i \omega)] X^\pr(u)\nonumber \ee \be +[5 + 4 a + 4 \sqrt{1+4 a}+4 \xi^2 -8 i \omega-4 i \sqrt{1+4 a} \omega -4 \omega^2]X(u)=0~,\ee
which may be readily solved in terms of hypergeometric functions
$$ X(u) = C_1 \ {_2F_1}\left(\f{1}{2}+\f{\sqrt{1+4 a}}{4}-\f{i \xi}{2} -\f{i\omega}{2},\ \f{1}{2}+\f{\sqrt{1+4 a}}{4}+\f{i \xi}{2} -\f{i\omega}{2}, 1 + \f{\sqrt{1+4 a}}{2},\ u\right)$$
\be +C_2 \ u^{-\f{\sqrt{1+4 a}}{2}} {_2F_1}\left(\f{1}{2}-\f{\sqrt{1+4 a}}{4}-\f{i \xi}{2} -\f{i\omega}{2},\ \f{1}{2}-\f{\sqrt{1+4 a}}{4}+\f{i \xi}{2} -\f{i\omega}{2}, 1 - \f{\sqrt{1+4 a}}{2},\ u\right).\ee Thus one obtains the solution of the original equation
$$ \Psi(u) $$ $$= C_1 \ u^{\f{1+\sqrt{1+4 a}}{4}} (1-u)^{-\f{i \omega}{2}}\ {_2F_1}\left(\f{1}{2}+\f{\sqrt{1+4 a}}{4}-\f{i \xi}{2} -\f{i\omega}{2},\ \f{1}{2}+\f{\sqrt{1+4 a}}{4}+\f{i \xi}{2} -\f{i\omega}{2}, 1 + \f{\sqrt{1+4 a}}{2},\ u\right)$$
\be +C_2 \ u^{\f{1-\sqrt{1+4 a}}{4}} (1-u)^{-\f{i \omega}{2}} {_2F_1}\left(\f{1}{2}-\f{\sqrt{1+4 a}}{4}-\f{i \xi}{2} -\f{i\omega}{2},\ \f{1}{2}-\f{\sqrt{1+4 a}}{4}+\f{i \xi}{2} -\f{i\omega}{2}, 1 - \f{\sqrt{1+4 a}}{2},\ u\right)~.\ee
In view of the above expressions when $1+4 a=0$ we get a critical value for the GB coupling $$\lambda_c=\sqrt{\f{3}{8}}\approx 0.61~.$$
%\subsubsection{Small Gauss--Bonnet coupling}

If $\lambda$ is small enough, i.e. $\lambda<\lambda_c,$ one should set $C_2=0$ to ensure finiteness at $u\to 0.$ The solution reduces to
\be \Psi(u) = C_1 \ u^{\f{1+\sqrt{1 + 4 a}}{4}} (1-u)^{-\f{i \omega}{2}}\ \ee\be \times{_2F_1}\left(\f{1}{2}+\f{\sqrt{1+4 a}}{4}-\f{i \xi}{2} -\f{i\omega}{2},\ \f{1}{2}+\f{\sqrt{1+4 a}}{4}+\f{i \xi}{2} -\f{i\omega}{2}, 1 + \f{\sqrt{1+4 a}}{2},\ u\right)~.\ee
The expansion of the hypergeometric function around $u=1$ reads
\be{_2F_1}\left(\f{1}{2}+\f{\sqrt{1+4 a}}{4}-\f{i \xi}{2} -\f{i\omega}{2},\ \f{1}{2}+\f{\sqrt{1+4 a}}{4}+\f{i \xi}{2} -\f{i\omega}{2}, 1 + \f{\sqrt{1+4 a}}{2},\ u\right) \ee\be \simeq K_1 \f{1}{\Gamma\left(\f{1}{2}+\f{\sqrt{1+4 a}}{4}-\f{i \xi}{2}+\f{i \omega}{2}\right) \Gamma\left(\f{1}{2}+\f{\sqrt{1+4 a}}{4}+\f{i \xi}{2}+\f{i \omega}{2}\right)\Gamma\left(1-i \omega \right)} \ee
\be + K_2 \f{1}{\Gamma\left(\f{1}{2}+\f{\sqrt{1+4 a}}{4}-\f{i \xi}{2}-\f{i \omega}{2}\right) \Gamma\left(\f{1}{2}+\f{\sqrt{1+4 a}}{4}+\f{i \xi}{2}-\f{i \omega}{2}\right)\Gamma\left(1+i \omega \right)} (1-u)^{+i \omega}~,\ee
 where $K_1$ and $K_2$ are constants, in the sense that they do not involve $u.$ Since we insist on having only terms of the form $(1-u)^{-\f{i \omega}{2}}$ near $u=1,$ it is obvious that the second term, which involves  $(1-u)^{+i \omega},$ should be discarded; the only way to discard it is the divergence to infinity of the $\Gamma$ functions in the denominator, which happens when
 \be\f{1}{2}+\f{\sqrt{1+4 a}}{4}+\f{i \xi}{2}-\f{i \omega}{2}=-n, \ \Rightarrow \omega =\pm\xi - i\left(2 n +\f{2+\sqrt{1+4 a}}{2}\right)\label{omegan}~.\ee
 The quantity $n$ is a non-negative integer. Thus we have determined (to zeroth approximation) the QNMs
  \be \omega_n=\pm\xi-i\left(2 n +\f{2+\sqrt{9-24\lambda^2}}{2}\right),\ \ n=0,1,2,\dots\ee
for small GB coupling $\lambda$, less than its critical value $\lambda_c$.
%\subsubsection{Large Gauss--Bonnet coupling}

If $\lambda$ grows enough, so that  $\lambda>\lambda_c,$ we work along similar lines and we get
\be \omega_n=\pm\xi+\f{\sqrt{24 \lambda^2-9}}{2}-i\left(2 n +1\right)~.\ee
Notice that the real part of the QNMs may be non-zero even when $\xi$ vanishes.

\subsubsection{Analytical predictions for the QNMs}

The above analysis holds strictly at $r_+=1.$ Thus we have a prediction for the results if $r_+=1:$
\begin{itemize}
  \item For small $\lambda$ we expect to find QNMs $\omega\equiv \omega_R-i \omega_I$ with a constant $\omega_R=\pm \xi$ (the same for all of them) and with $\omega_I,$ with an interval $2$ between successive values.
  \item For large $\lambda$ we expect $\omega_R = \pm\xi+\f{\sqrt{24\lambda^2-9}}{2}.$
\end{itemize}

\begin{figure}[ht]
\includegraphics[width=8cm]{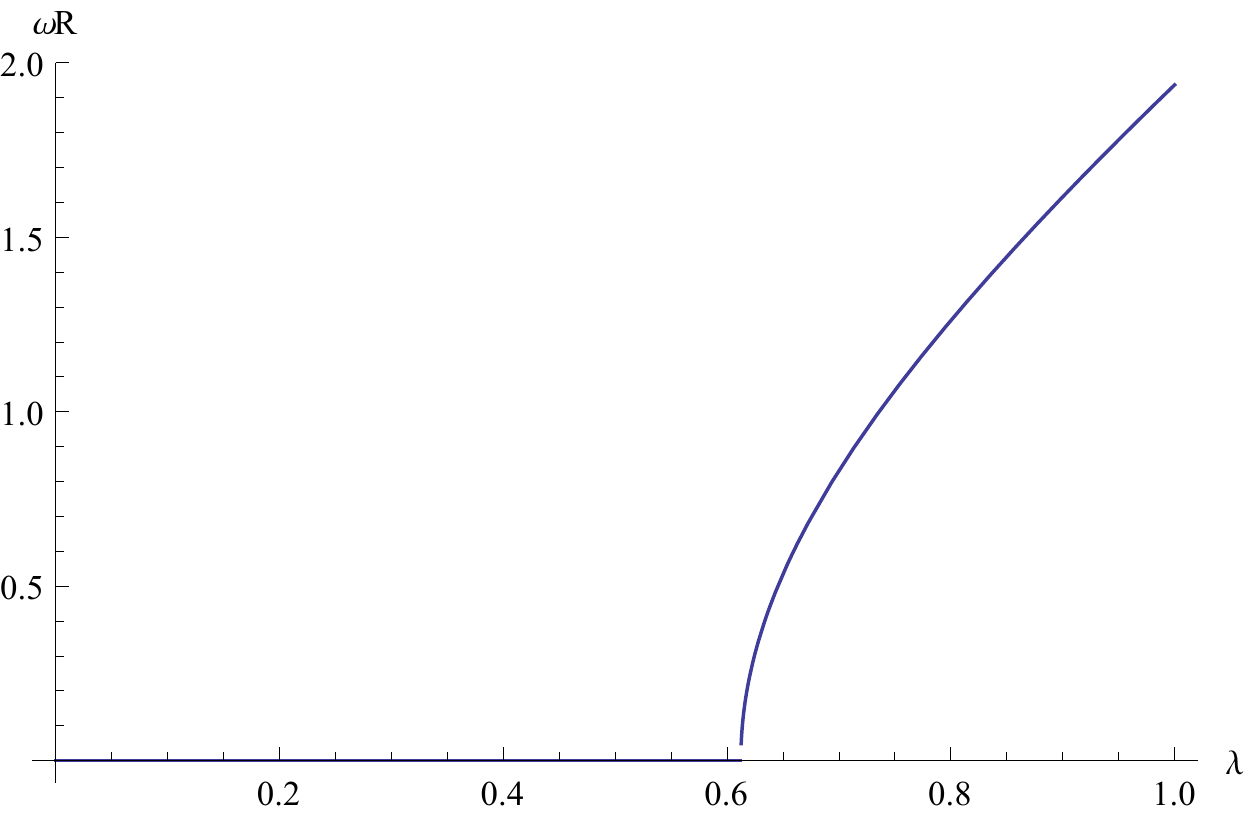}
\includegraphics[width=8cm]{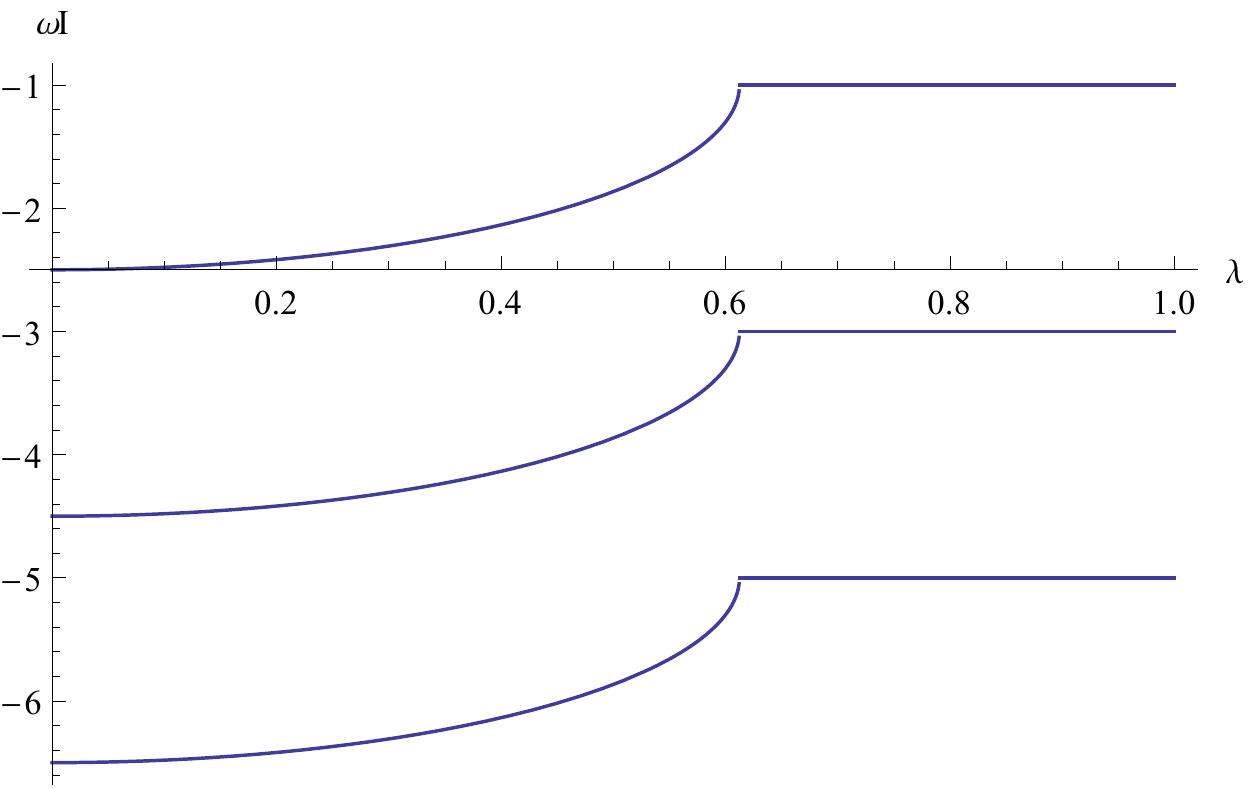}
\centering
\caption{$\omega_R$ (left) and $\omega_I$ (right) versus $\lambda$ for $\xi=0$ and $n=0,\ 1$ and $2.$ The value $n=0$ corresponds to the uppermost curve.}
\centering
\label{omegas}
\end{figure}

One may depict the above changes in the Fig.~\ref{omegas}, where the quantities $\omega_R$ and $\omega_I$ are shown versus $\lambda.$ It is evident that a qualitative change happens at $\lambda=\lambda_c,$ since the slope presents a discontinuity. It is reasonable to expect a phase transition to happen at this value of $\lambda.$ The real part, $\omega_R,$ vanishes for small $\lambda,$ that is $\lambda<\lambda_c,$ while it takes non-zero values for large $\lambda,$ even though $\xi$ is set to zero. In addition, it does not depend on the integer $n.$ On the other hand, the imaginary part, $\omega_I,$ depends on the integer $n.$

In the numerical calculations of the QNMs one works actually for values for $r_+$ either slightly smaller or slightly larger than $1.$ In
\cite{Koutsoumbas:2008yq} it was  found that the real part for $r_+\ne 1$ is no more the same for all QNMs, as predicted above, there is a slope, which is positive (negative) for $r_+<1\ (r_+>1).$ Thus the QNMs are infinite in number for $r_+>1,$ while for $r_+<1$ the QNMs approach the vertical axis and eventually cease to exist. In addition, if $\xi$ is small enough, no propagating modes exist.

From Fig.~\ref{omegas} we can see that for $\lambda<\lambda_c,$ the real part of the QNMs is zero while the imaginary part is negative. These results indicate that for values of $\lambda$ below its critical value the TBH is stable under scalar perturbations, while above that critical value of $\lambda$ we have instability indicating that we have a phase transition of the TBH to a MTZ-like black hole.

If we want to go to different (but close enough) values of $r_+,$ we may calculate corrections analytically, using perturbation theory. However this is technically difficult, so we postpone it for a future work. We expect that instabilities may show up in a perturbative calculation for $r_+\ne 1,$ when $\lambda$ takes on sufficiently large values.

\subsection{Large black holes}

Next we examine the limit $r_+\to +\infty.$ In our case we have $\hat{g}(u)=\f{1}{u}-\f{1}{r_+^2}-\f{M}{r_+^3}\sqrt{u}$ and the condition $\hat{g}(1)=0$ yields $M=r_+^3-r_+\simeq r_+^3,$ since $r_+$ is large.
Taking this into account we find that $u^{3/2}\hat{g} =\sqrt{u}-u^2.$ In addition $$\hat{{\cal V}} =\hat{g}\left[2+u^{3/2}-6 \lambda^2-3\lambda^2 u^3\right]~.$$
The Klein-Gordon equation reads
\be -4 u^{3/2}\hat{g}\left((\sqrt{u}-u^2) \Psi^\pr\right)^\pr + \hat{g}\left[2+u^{3/2} - 6 \lambda^2-3\lambda^2 u^3\right]\Psi=\f{\omega^2}{r_+^2}\Psi~.\ee It is natural to assume that
$\f{\omega^2}{r_+^2}\to 0,$ as $r_+\to +\infty,$
so Klein-Gordon  equation simplifies to
\be -4 u^{3/2}\left((\sqrt{u}-u^2) \Psi^\pr\right)^\pr + \left[2+u^{3/2}-6 \lambda^2-6\lambda^2 u^3\right]\Psi=0~.\ee
We neglect the $u^3$ term and we set $a \equiv 2 -6 \lambda^2,$ so equation the
\be -4 u^{3/2}\left((\sqrt{u}-u^2) \Psi^\pr\right)^\pr + \left[a+u^{3/2}\right]\Psi=0\ee
 is solved by
  $$\Psi=C_1 u^{\f{1-\sqrt{1 +4 a}}{4}} \ _2F_1\left(\f{1}{2}-\f{\sqrt{1 +4 a}}{6},\f{1}{2}-\f{\sqrt{1 +4 a}}{6},1-\f{\sqrt{1 +4 a}}{3},u^{3/2}\right)$$ $$+ C_2 u^{\f{1+\sqrt{1 +4 a}}{4}} \ _2F_1\left(\f{1}{2}+\f{\sqrt{1 +4 a}}{6},\f{1}{2}+\f{\sqrt{1 +4 a}}{6},1+\f{\sqrt{1 +4 a}}{3},u^{3/2}\right)~.$$ For $\lambda=0$
  the solution reduces to \be \Psi=C_1 \f{\ln(1-u^{3/2})}{\sqrt{u}}+C_2 \f{1}{\sqrt{u}}~,\ee
  which diverges at both $u=0$ and $u=1.$
  %This is an impasse for this model, unless one supposes that $\f{\omega^2}{r_+^2}$ does not vanish in the limit of large $r_+,$ which results in %unphysical QNMs, which are proportional to $r_+.$
  In the case that $\lambda\ne 0$ the limits are the following
   \be \lim_{u\to 0} \Psi  =C_1 u^{\f{1-\sqrt{1 +4 a}}{4}} + C_2 u^{\f{1+\sqrt{1 +4 a}}{4}}~,\ee which may be finite for a wise choice of the integration constants.
   However
\be \lim_{u\to 1} \Psi=C_1 u^{\f{1-\sqrt{1 +4 a}}{4}} \ _2F_1\left(\f{1}{2}-\f{\sqrt{1 +4 a}}{6},\f{1}{2}-\f{\sqrt{1 +4 a}}{6},1-\f{\sqrt{1 +4 a}}{3},1\right)$$ $$+ C_2 u^{\f{1+\sqrt{1 +4 a}}{4}} \ _2F_1\left(\f{1}{2}+\f{\sqrt{1 +4 a}}{6},\f{1}{2}+\f{\sqrt{1 +4 a}}{6},1+\f{\sqrt{1 +4 a}}{3},1\right)~.\ee
We recall that
$$ _2F_1(A,B,C,1)=\f{\Gamma(C)\Gamma(C-A-B)}{\Gamma(C-A)\Gamma(C-B)}~.$$
 Setting $A,\ B,\ C$ equal to the relevant values we find
$$_2F_1\left(\f{1}{2}-\f{\sqrt{1 +4 a}}{6},\f{1}{2}-\f{\sqrt{1 +4 a}}{6},1-\f{\sqrt{1 +4 a}}{3},1\right) $$
$$= \f{\Gamma(1+\f{\sqrt{1 +4 a}}{3})\Gamma(0)}{\Gamma(\f{1}{2}+\f{\sqrt{1 +4 a}}{6})\Gamma(\f{1}{2}+\f{\sqrt{1 +4 a}}{6})}~.$$
These relations, with the infinite expression $\Gamma(0)$ in their right hand sides, seem to indicate that, also in this case, only unphysical QNMs with $\lim_{r_+\to +\infty}\f{\omega^2}{r_+^2}\ne 0$ may exist.

We have not been able to analytically investigate the regime of tiny black holes, that is the ones with $r_+<<1.$ However we have used the method described in the Section of scalar perturbations and it seems that there are no QNMs for tiny black holes. Thus it is plausible that only horizons around one may be expected to yield QNMs.

\subsection{Scalar modes}

\begin{figure}[ht]
\includegraphics[width=8cm]{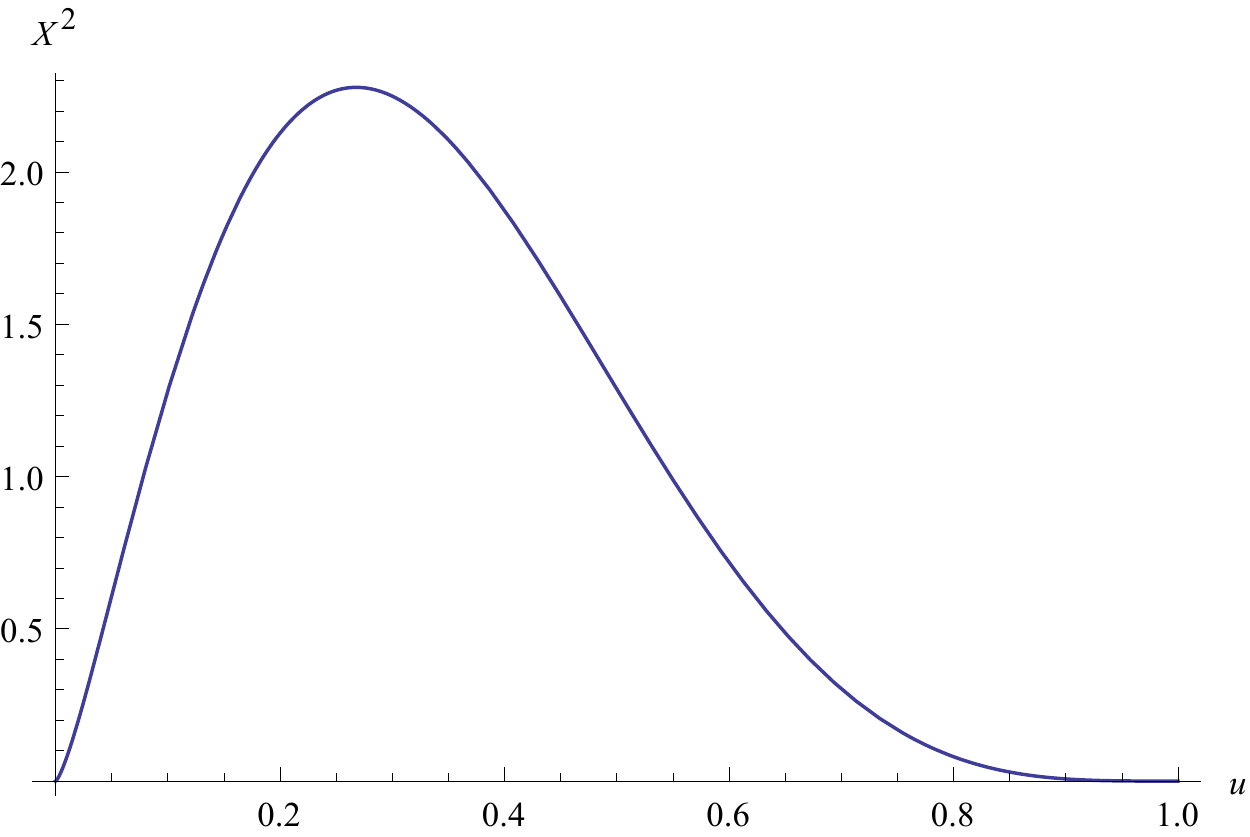}
\centering
\caption{$X^*(u) X(u)$ for the scalar field versus $u$ for $\lambda=0.5$ and $\xi=0.$}
\centering
\label{lambda05}
\end{figure}

To calculate the scalar modes we start with the equation (\ref{eqhor}) when the horizon equals one
 \be -4 u^{1/2} (u^{1/2} (1-u) \Psi^\pr)^\pr +\left(\f{2 - 6 \lambda^2}{u}+\xi^2+\f{1}{4}\right)\Psi=\f{\hat{\omega}_n^2}{1-u}\Psi~.\ee
 We employ the transformation (\ref{transf})
  \be \Psi(u)=u^{\f{1+\sqrt{9-24\lambda^2}}{4}} (1-u)^{-\f{i \omega_n}{2}} X(u)~,\ee
and use the result (\ref{omegan})
 \be\omega_n= \pm \xi - i\left(2 n +\f{2+\sqrt{9-24\lambda^2}}{2}\right)~.\ee
 The resulting equation reads
 \be 2(-1+u) u X^\pp + [2 + \sqrt{9-24 \lambda^2}+u (-2+4 n+2 i \xi)] X^\pr + 2 n (n+i\xi) X=0~.\ee
 Let us check the behaviour of the quantity
 \be \hat{\rho}=\Psi^*(u) \Psi(u)~,\ee
 which contains the u-dependence of the charge density
\be \rho=\f{e}{2 m}\left[\left(\Psi e^{-i \omega t}\right)^*\p_t \left(\Psi e^{-i \omega t}\right) - \left(\Psi e^{-i \omega t}\right)\p_t \left(\Psi e^{-i \omega t}\right)^*\right]_{t=0}~.\ee
For $n=0$ the equation simplifies to
 \be 2(-1+u) u X^\pp + [2+\sqrt{9-24 \lambda^2}+u (-2+2 i \xi)] X^\pr = 0~.\ee
 The solution reads
 \be X=c_1+c_2 u^{-\f{1}{2}\sqrt{9-24\lambda^2}}\ _2F_1\left(-\f{1}{2}\sqrt{9-24\lambda^2},\ -\f{1}{2}\sqrt{9-24\lambda^2}-i \xi,\ -\f{1}{2}\sqrt{9-24\lambda^2},\ u\right)~.\ee
 However the factor $u^{-\f{1}{2}\sqrt{9-24\lambda^2}}$  does not behave properly in the limit $u\to 0,$ if $\lambda<\lambda_c,$ so, in this case, the result is a constant function, that is, $c_2$ must be set to zero. Thus the solution for $\lambda<\lambda_c$ takes the form
\be \Psi_<(u)=c_1 u^{\f{1+\sqrt{9-24 \lambda^2}}{4}} (1-u)^{-\f{i \omega}{2}},\ \  \omega=\pm\xi-i\f{2+\sqrt{9-24 \lambda^2}}{2}~.\ee
Its solution for $\lambda>\lambda_c$ is the linear combination
\be \Psi_>(u)=  u^{\f{1}{4}}u^{\f{i \sqrt{24 \lambda^2-9}}{4}} (1-u)^{-\f{i \omega_n}{2}} \ee\be\times\left[c_1+c_2 u^{-\f{i}{2}\sqrt{24\lambda^2-9}}\ _2F_1\left(-\f{i}{2}\sqrt{24\lambda^2-9},\ -\f{i}{2}\sqrt{24\lambda^2-9}-i \xi,\ -\f{i}{2}\sqrt{24\lambda^2-9},\ u\right)\right]~, \ee
\be \ \omega_n=\pm\left(\xi+\f{\sqrt{24 \lambda^2-9}}{2}\right)-i\left(2 n +1\right)~.\ee
We set $\lambda=0.5,$ which lies in the region of small $\lambda$ values. We may check by inspection that there is no dependence on $\xi$ in this region of $\lambda$ (and in this approximation). Fig.~\ref{lambda05} displays the result.

\begin{figure}[ht]
\includegraphics[width=7cm]{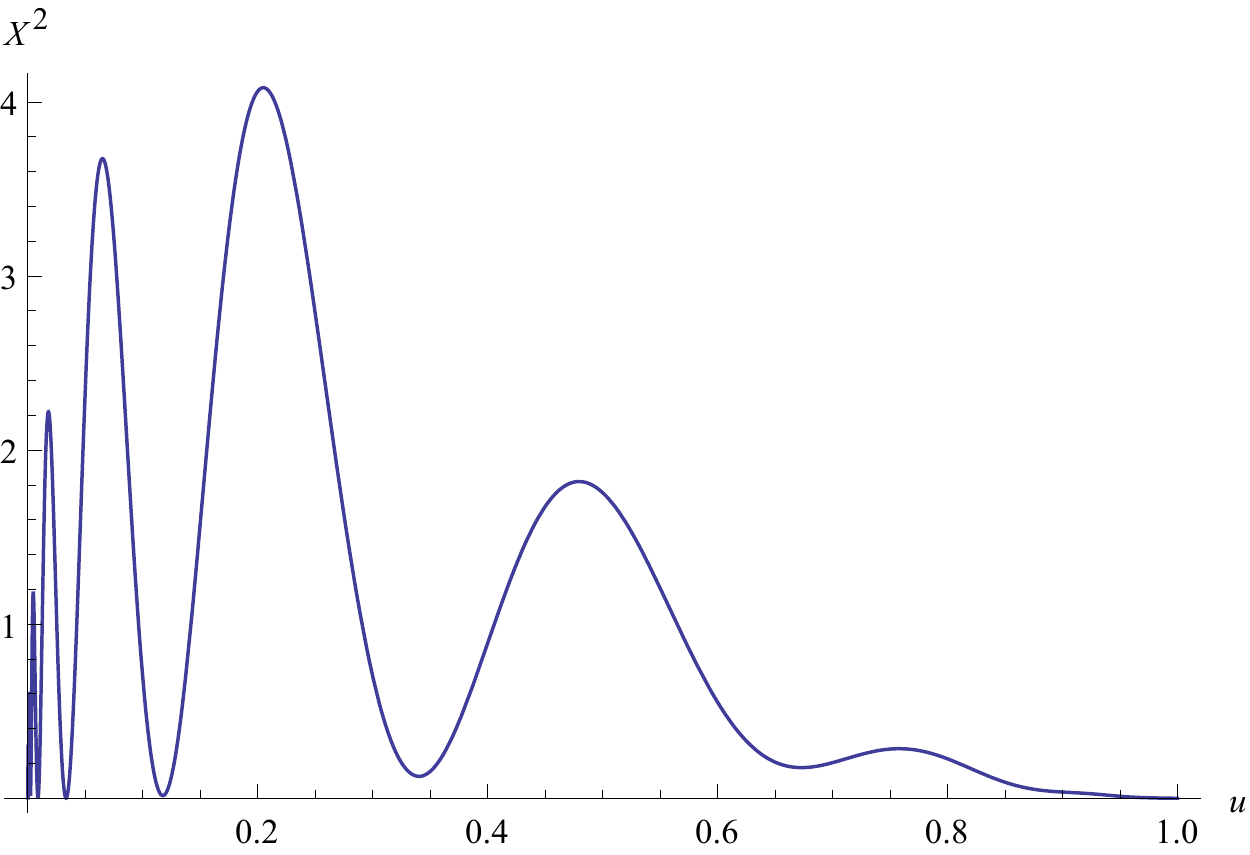}
\includegraphics[width=7cm]{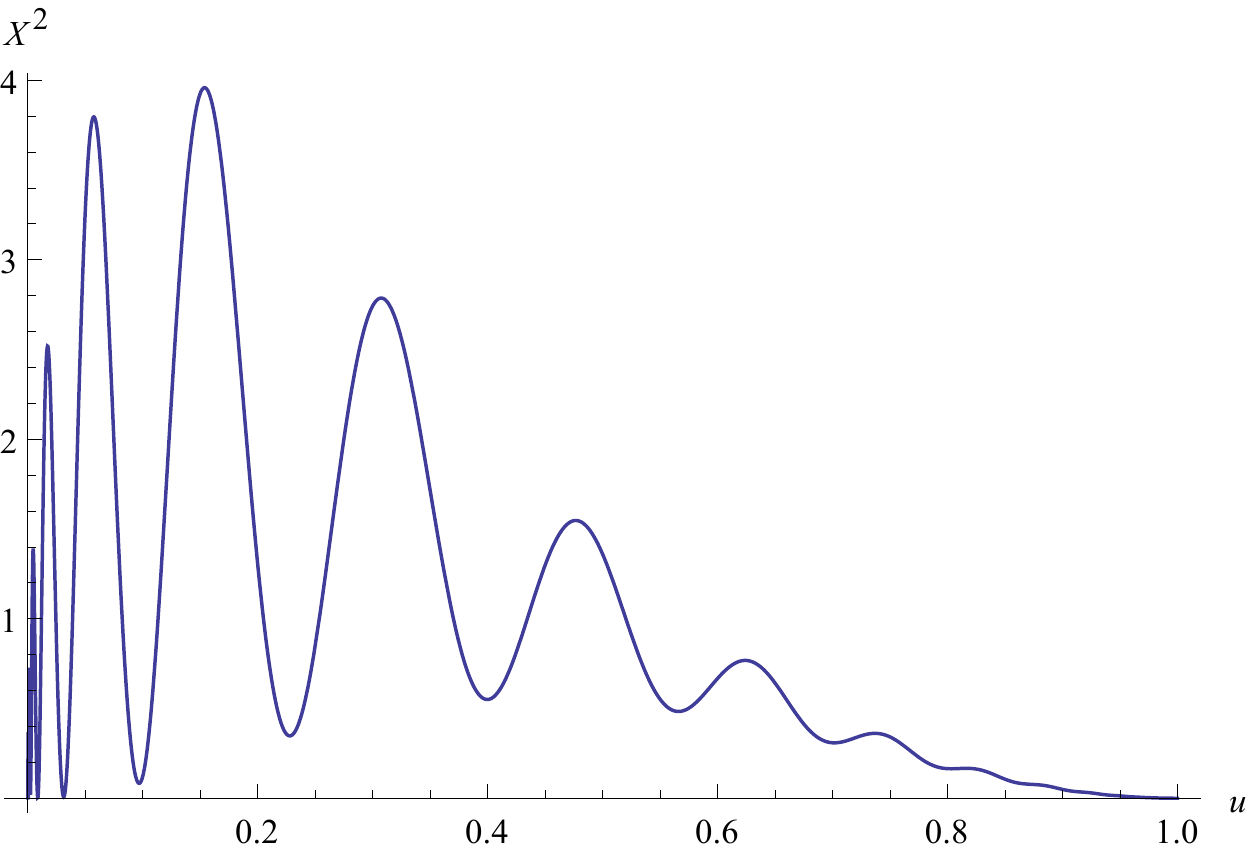}
\centering
\caption{$X^*(u) X(u)$ for the scalar field versus $u$ for $\lambda=2.0$ and either $\xi=0$ (left plot) or $\xi=10$ (right plot).}
\centering
\label{lambda20}
\end{figure}

In Fig.~\ref{lambda20} one may observe the results for $\lambda=2.0,$ that is $\lambda>\lambda_c,$ at $\xi=0$ and $\xi=10.$ The most striking characteristic is the qualitative difference between Fig.~\ref{lambda05} and Fig.~\ref{lambda20} (a), which may lend support to the conjecture that, moving to large values of $\lambda,$ may result in a phase transition. When one uses $\xi=10,$ quantitative differences are evident, in contrast to the previous case, but these differences do not qualify for a qualitative change.

%%%%%%%%%%%%%%%%%%%%%%%%%%%%%%%%%%%%%%%%%%%%5George%%%%%%%%%%%%%%%%%%%%%%%%%%%%%%%%%%%%%%%%%%%%%%%%%%%%

\section{Scalar Perturbations}
\label{sec4}

In this Section we will consider scalar perturbation in ESTGB gravity theories in the case where the background metric is the TBH. In the case of a trivial scalar field the  equation which describes massive scalar perturbations in this spacetime background reads
\begin{equation}
    \left( \Box_{(0)}  - \frac{m^2}{2} +\frac{1}{4}\lambda^2  \mathcal{R}^2_{GB(0)}\right)\delta\phi=0~, \label{scalarperturbationsTBH}
\end{equation}
where $\Box_{(0)}$ and $\mathcal{R}^2_{GB(0)}$ are the D'Alembert operator and the Gauss-Bonnet invariant for the topological geometry. So
\begin{equation}
\mathcal{R}^2_{GB(0)}=\frac{4 \left((g(r)+1) g''(r)+g'(r)^2\right)}{r^2}=\frac{24}{L^4}+\frac{12 M^2}{r^6}~. \label{gbform}
\end{equation}
This small perturbation has the same symmetries of the TBH, namely static and spherical symmetry. So the variables can be decomposed by the standard way
\begin{equation}
    \delta\phi= u(r) e^{-i \omega t}{\cal Y}_{\xi m}(\theta,\varphi)~.
\end{equation}
Note that the spherical harmonics ${\cal Y}_{\xi m}(\theta,\varphi)$ obey the equation
\begin{equation}
    \frac{1}{\sinh{\theta}}\partial_\theta\left( \sinh{\theta} \partial_\theta {\cal Y}_{\xi m}(\theta,\varphi)\right) + \frac{1}{\sinh^2{\theta}} \partial^2_\phi {\cal Y}_{\xi m}(\theta,\varphi) = -\left(\xi^2+\frac{1}{4} \right) {\cal Y}_{\xi m}(\theta,\varphi)~.
\end{equation}
After substituting in Eq.~(\ref{scalarperturbationsTBH}) and introducing the tortoise coordinate $dr^*=\dfrac{1}{g(r)}dr$, we obtain the following Schrödinger-like
equation
\begin{equation}
    u''(r^*)+\left(\omega^2-\mathcal{U}(r) \right) u(r^*)=0~, \label{SCHlikescalarequation}
\end{equation}
where the effective potential $\mathcal{U}(r)$ reads
\begin{equation}
    \mathcal{U}(r)=g(r) \left( \frac{1}{2}m^2+\frac{1}{r}g'(r)
    -\frac{\lambda^2}{4} \mathcal{R}^2_{GB(0)}+ \frac{\xi^2+\frac{1}{4}}{r^2} \right)~,
\end{equation}
and using (\ref{gbform}) we have
\begin{align}
    \mathcal{U}(r)=\left(-1-\frac{M}{r}+\frac{r^2}{L^2}\right) &\left( \frac{1}{2}m^2+\frac{1}{r}\left(\frac{M}{r^2}+\frac{2r}{L^2}\right) \right.\nonumber\\
    &\left.-\frac{\lambda^2}{4} \left(\frac{24}{L^4}+\frac{12M^2}{r^6} \right)+ \frac{\xi^2+\frac{1}{4}}{r^2} \right)~.
\end{align}
In the case of a non-trivial scalar field ({\ref{Psi}) the wave equation reads
 \be -\f{1}{\sqrt{-g}}\p_\mu[\sqrt{-g} g^{\mu\nu} \p_\nu\Psi]+\f{d U}{d \Psi}=0~.\ee
In the TBH background we have
$$ \f{1}{\sqrt{-g}}\p_\mu[\sqrt{-g} g^{\mu\nu} \p_\nu\Psi] =  -\f{1}{g(r)} \p_{tt}\Psi+\f{1}{r^2}\p_r[r^2 g(r) \p_r\Psi] +\f{1}{r^2}\f{1}{\sinh\theta}\p_\theta[\sinh\theta \p_\theta\Psi] +\f{1}{r^2\sinh^2\theta}\p^2_\phi\Psi~.$$
On the other hand for the spherical harmonics ${\cal{Y}}^{(k)}_q$ we have
 \be \f{1}{\sinh\theta}\p_\theta[\sinh\theta \p_\theta {\cal{Y}}^{(k)}_q] +\f{1}{\sinh^2\theta}\p^2_\phi{\cal{Y}}^{(k)}_q  = -\left(\xi^2+\f{1}{4}\right){\cal{Y}}^{(k)}_q~,\ee
 while the potential reads
  \be U=\f{1}{2} m^2 \Psi^2 -\f{\lambda^2}{2} f(\Psi) R_{GB}^2~,\ee
  where
   \be R_{GB}^2=R^2-4 R_{\mu\nu}R^{\mu\nu}+R_{\mu\nu\alpha\beta} R^{\mu\nu\alpha\beta} \rightarrow \f{4}{r^2}[g^{\pr 2}(r) +(g(r)+1) g^\pp(r)]~,\ee
   so that, replacing $\Psi(t,r,\theta,\phi) =\Phi(t,r) {\cal{Y}}^{(k)}_q (\theta,\ \phi)$ one ends up with an equation for a field depending just on $t$ and $r$,
\be \f{1}{g(r)} \p_{tt}\Phi(t,r)-\f{1}{r^2}\p_r[r^2 g(r) \p_r\Phi(t,r)] +\f{\xi^2+\f{1}{4}}{r^2}\Phi+m^2\Phi(t,r)-\lambda^2 R_{GB}^2 \f{d F}{d\Phi} = 0~.\ee

If we fix the scalar function to $F=\f{1}{2} \Phi^2,$ the scalar field equation becomes
 \be \f{1}{g(r)} \p_{tt}\Phi(t,r)-\f{1}{r^2}\p_r[r^2 g(r) \p_r\Phi(t,r)] +\f{\xi^2+\f{1}{4}}{r^2}\Phi+m^2\Phi(t,r)-\lambda^2 R_{GB}^2 \Phi = 0~.\ee

In this scalar field equation there is a direct coupling of the scalar field to the GB term and also an extra parameter  $\xi$  appears because of the hyperbolic geometry.

Changing the variables to
\be \Phi(t,r)=\f{\chi(t,r)}{r}\Rightarrow r^2 \p_r \Phi= r \chi^\pr-\chi\Rightarrow \p_r(r^2 \p_r \Phi) = r \chi^\pp~,\ee
the scalar equation becomes
\be \p_{tt}\chi - g(r) \f{d}{dr}\left(g(r) \f{d}{dr}\chi\right) + \f{g(r)}{r} \f{d f(r)}{dr} \chi + g(r) \f{\xi^2+\f{1}{4}}{r^2}\chi+m^2 g(r) \chi-\lambda^2 g(r) R_{GB}^2 \chi=0~.\label{ini}\ee
Introducing tortoise coordinates \be dr_*=\f{dr}{g(r)}\Leftrightarrow g(r)\f{d}{dr}=\f{d}{d r_*}~,\ee the equation takes the form
\be \p_{tt}\chi- \f{d^2}{dr_*^2}\chi +g(r) \left[\f{1}{r} \f{d g(r)}{dr} + \f{\xi^2+\f{1}{4}}{r^2}+m^2 -\lambda^2 R_{GB}^2 \right]\chi=0~.\ee
The time dependence of $\chi$ is $e^{-i \omega t}$ and the above equation takes the Schrödinger-like
form
\be -\f{d^2}{dr_*^2}\chi +g(r) V(r)\chi=\omega^2 \chi~,\ee
where the potential is given by
\be V(r) \equiv \f{1}{r}\f{d g(r)}{dr} + \f{\xi^2+\f{1}{4}}{r^2}+m^2 -\lambda^2 R_{GB}^2~. \ee
For the TBH the potential becomes
\be V(r) = \f{2}{L^2}+\f{2 M}{r^3} + \f{\xi^2+\f{1}{4}}{r^2}+m^2 -\lambda^2 R_{GB}^2~. \label{pot33} \ee
Setting
$$\chi= \psi_\omega e^{-i\omega r_*}~,$$
the scalar field equation becomes
\be g(r) \f{d^2\psi_\omega}{dr^2} +\left(\f{dg(r)}{dr}-2 i\omega\right) \f{d\psi_\omega}{dr}=V(r) \psi_\omega~,\label{eq2a}\ee
 where $V(r)$ is given by (\ref{pot33}).
 To   investigate the properties of the scalar field it is useful to change variables from $r$ to $x=\f{1}{r}.$ Then
equation (\ref{eq2a}) is transformed into
   \be h(x) \left[x^4 \f{d^2 \psi_\omega}{d x ^2}+2 x^3 \f{d \psi_\omega}{d x}\right] +\left(-x^2 \f{d h(x)}{d x}-2 i\omega\right) \left[-x^2 \f{d \psi_\omega}{dx}\right]=V(x) \psi_\omega,\ \ h(x) = \f{1}{L^2 x^2}-2 M x-1~, \label{eq3}\ee
    where
    \be V(x) = \f{2}{L^2}+2 M x^3 + \left(\xi^2+\f{1}{4}\right) x^2+ m^2 -24 \lambda^2 \left[\f{1}{L^4}+2 M^2 x^6\right]~.\ee
 The horizon variable $x_+$ is determined through
    \be h(x_+)=0\Rightarrow \f{1}{L^2 x_+^2}-2 M x_+-1=0\Rightarrow 2 M=\f{1}{L^2 x_+^3}-\f{1}{x_+}~.\ee
 This means that \be h(x) = \f{1}{L^2 x^2}-\left(\f{1}{L^2 x_+^3}-\f{1}{x_+}\right) x-1\Rightarrow \f{d h(x)}{dx}=-\f{2}{L^2 x^3}+\f{1}{x_+}~,\ee
so that the metric function and the potential take the form
$$h(x)=(x-x_+)\f{L^2 x^2 x_+^2-x^2-x_+^2-x x_+}{L^2 x_+^3 x^2}~,$$ and \be V(x) = \f{2}{L^2}+2 M x^3 + \left(\xi^2+\f{1}{4}\right) x^2+ m^2 -24 \lambda^2 \left[\f{1}{L^4}+\f{(1-x_+^2)^2}{2 x_+^6} x^6\right]~.\ee
The introduction of the horizon radius in the scalar field equation will be helpful to study the behaviour of the scalar field near and far away from the horizon of the black hole.

We expand $\psi_\omega$ about $x_+$
\be \psi_\omega(x)=\sum_k a_n(\omega) (x-x_+)^n~,\ee
and using the scalar field equation we find a recurrence formula of the form
\be a_n(\omega) = -\f{1}{P_{n,0}}\sum_{m=n-7}^{n-1} P_{m,n-m} a_m(\omega),\ \ P_{m,n-m}=m(m-1) s_{n-m}+m t_{n-m}+u_{n-m}~,\ee
where
 $$s(x)=\f{x^4 h(x)}{x-x_+},\,\,\, u(x)= -(x-x_+) V(x)~.$$
For the consistency of our calculations we  demand that the wave function vanishes at infinity $(r\to \infty, x=0),$ which yields the equation
\be \psi_\omega(0)=\sum_k a_n(\omega) (-x_+)^n =0~.\label{qnm}\ee
One has to solve the scalar field equation for $\omega,$ which are the quasi-normal frequencies. We solve the scalar equation numerically and we plot the points of the complex $\omega$ plane, where $\psi_\omega(0)$ vanishes. The method we use is to make a contour plot for each of the real and the imaginary part of $\psi_\omega(0),$ that is, find the points where each of the above vanishes. The points that we are looking for are exactly the points of intersection of the various curves. We have used between 500 and 1000 terms in the above sums, the criterion being the stabilization of the results.

\subsection{QNMs for $\lambda=0.5$ and $ \xi=0$}

As we saw in the analytic calculation, the system becomes unstable for large values of $\lambda,$ larger that about $0.61.$ At first we will consider values safely below this value.

We will consider $\lambda$ to take the value $0.5,$ where we do not expect instabilities. As can be seen in Fig.~\ref{fig1}, left panel,   the  intersections of the curves for $r_+=1.10$ lie in the negative $\omega_I$ half-plane, the line connecting them has a negative slope and the consecutive imaginary parts differ by $2 i$. As one considers larger black holes, that is, larger $r_+,$ the QNMs move towards less negative values: a relevant result is shown in Fig.~\ref{fig1}, right panel, where $r_+=0.95$ and $r_+=1.60.$; in addition the differences between consecutive QNMs increase in magnitude, that is, the QNMs appear more sparse. At some value of $r_+$ the intersections disappear completely, in agreement with our previous result that no QNMs exist for large black holes.

\begin{figure}[H]
\centering
\includegraphics[scale=0.4]{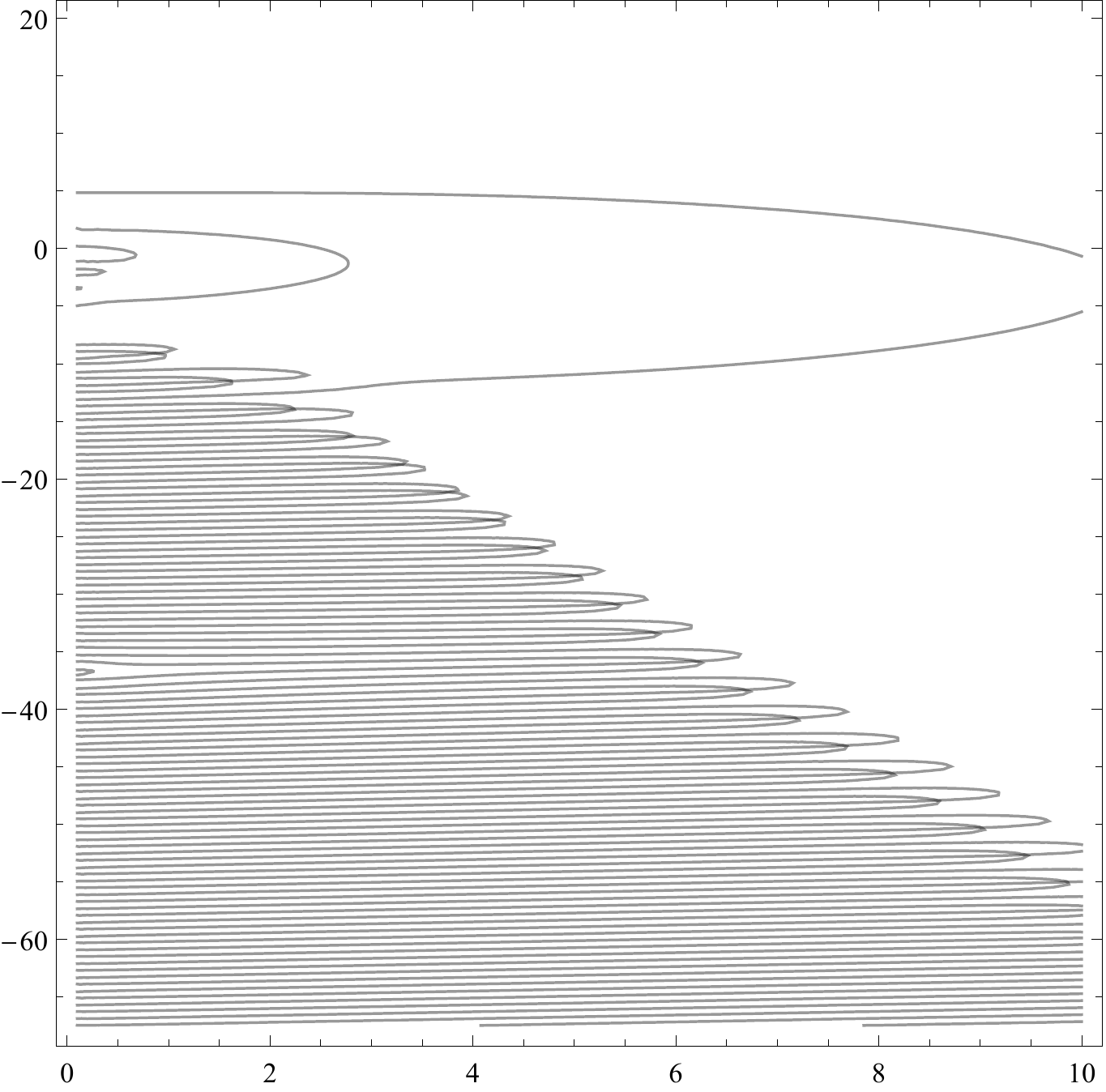}
\includegraphics[scale=0.4]{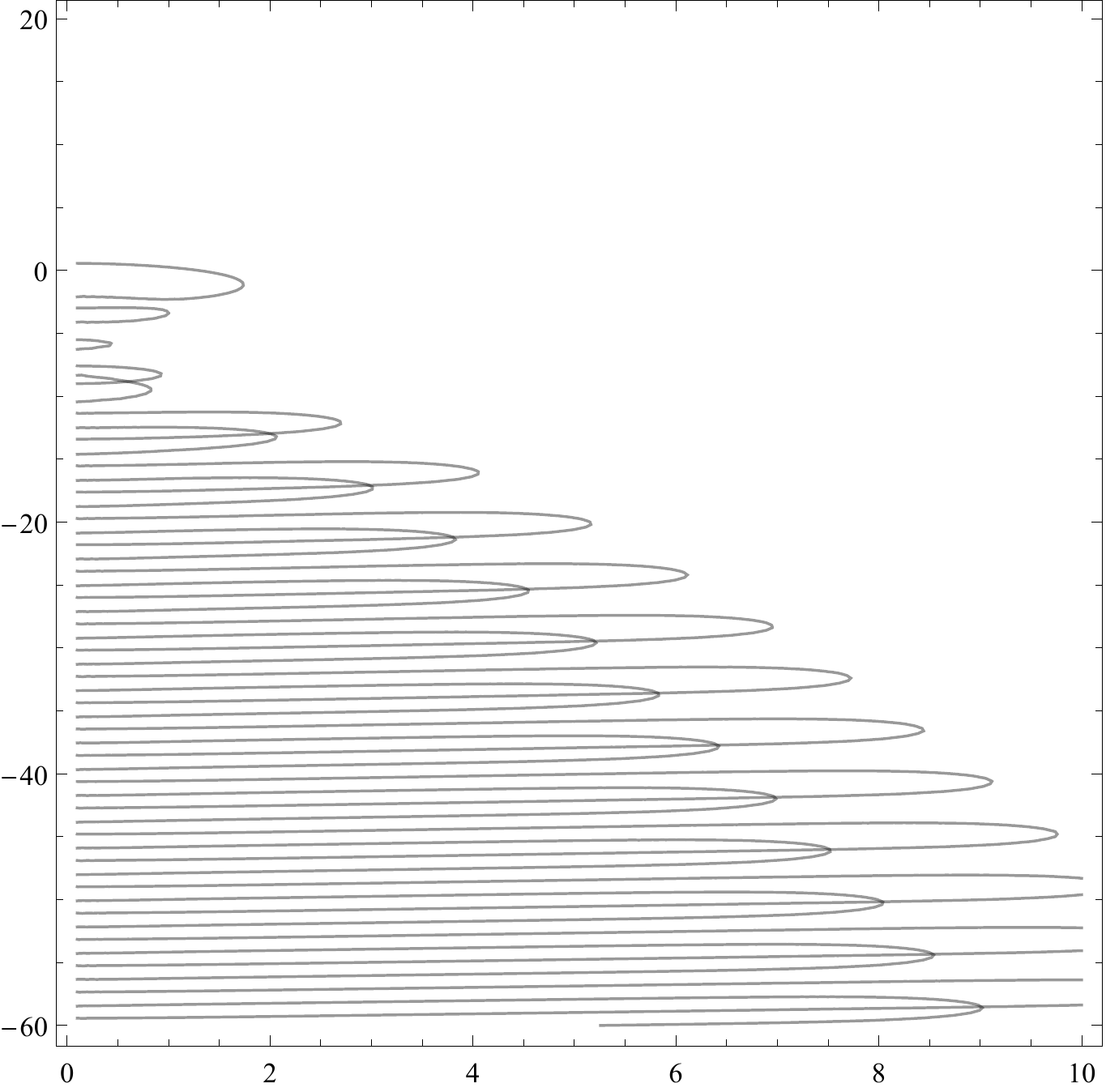}\\
\caption{$\lambda=0.5,\ \xi=0,\ r_+=1.10,\ r_+=1.60.$} \label{fig1}
\end{figure}

In Fig.~\ref{fig3}, $\lambda$ is set to the relatively large value $\lambda=1.5,$ $\xi$ is set to 0 and $r_+$ takes on the values $1.10$ and $2.00.$ For $r_+=1.10$ (left panel), apart from the QNMs with $\omega_I<0,$ {\bf there exist several QNMs with $\omega_I>0,$} signalling {\bf instability.} It is conceivable that this instability means that the metric used is no longer operational and {\bf scalarization} should be considered. We note that the QNMs with $\omega_I>0$ have a positive slope. The number of QNMs with $\omega_I>0$ decreases as $r_+$ increases, until at $r_+=2.00$ (right panel) they disappear completely. If we keep increasing $r_+,$ even the QNMs with $\omega_I<0$ disappear, in agreement with the analytical prediction that no QNMs exist for large black holes.

\begin{figure}[H]
\centering
\includegraphics[scale=0.4]{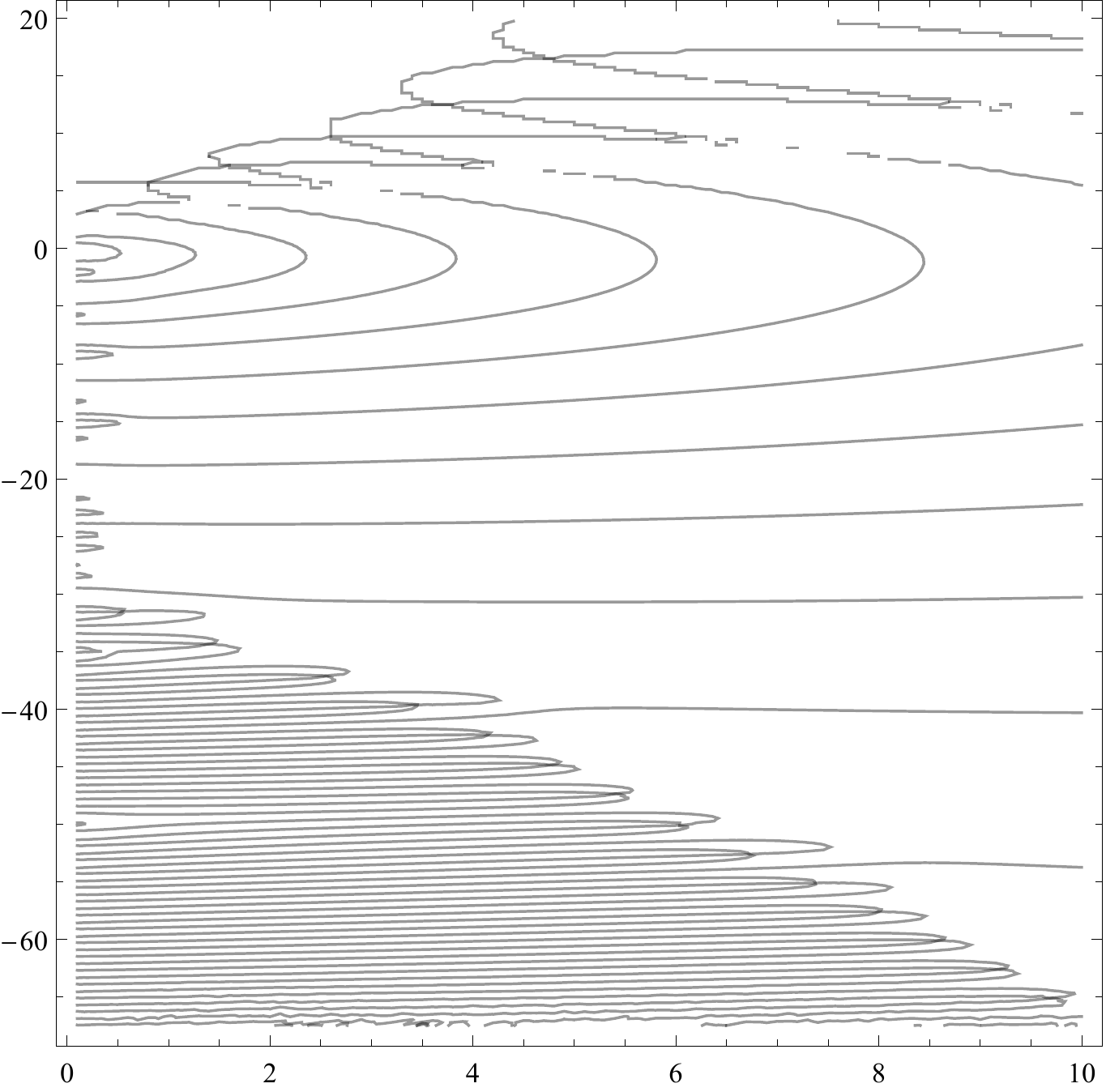}
\includegraphics[scale=0.44]{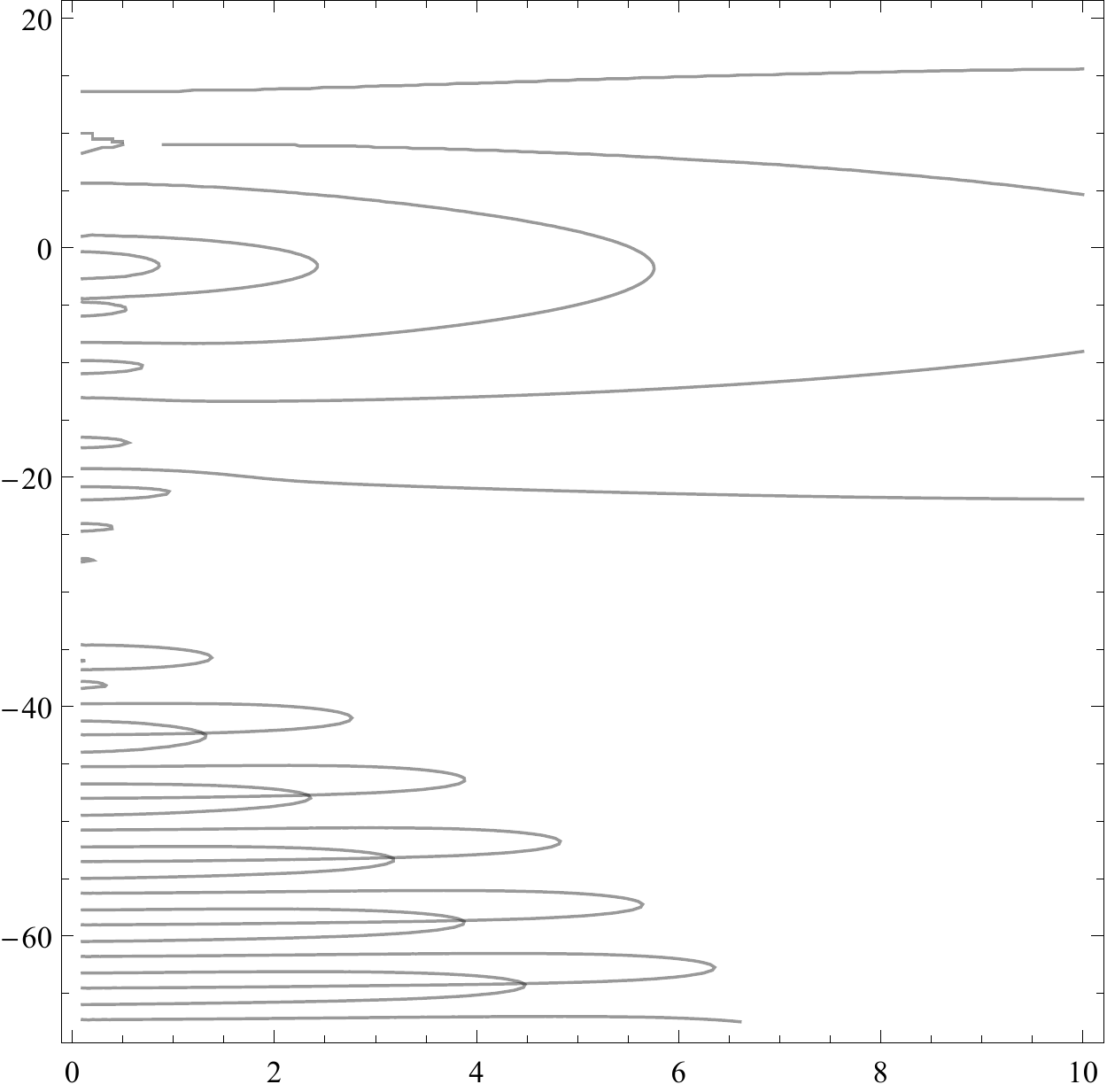}\\
\caption{$\lambda= 1.5,\ \xi=0.0,\ $ and $r_+=1.10,\ 2.00.$} \label{fig3}
\end{figure}

\subsection{QNMs for $\xi=0,\ r_+=1.10$ and large $\lambda.$}

Fig.~\ref{fig7} refers to the dependence of the QNM's on $\lambda,$ when $\xi=0$ and $r_+=1.10.$ For $\lambda=1.50$ (left panel) QNM's exist with negative $\omega_I.$ In addition QNM's with positive values of $\omega_I$ appear, whose existence gets more pronounced as $\lambda$ increases. For $\lambda=3.0,$ the QNM's with negative $\omega_I$ disappear. This picture persists for even larger values of $\lambda.$ This is consistent with the remark made earlier that the expression $\sqrt{9-24\lambda^2},$ appearing in the analytical treatment, suggests that, for large values of $\lambda,$ instabilities are expected to set in.

\begin{figure}[H]
\centering
\includegraphics[scale=0.4]{lambda15_xi0_rp110.pdf}
\includegraphics[scale=0.4]{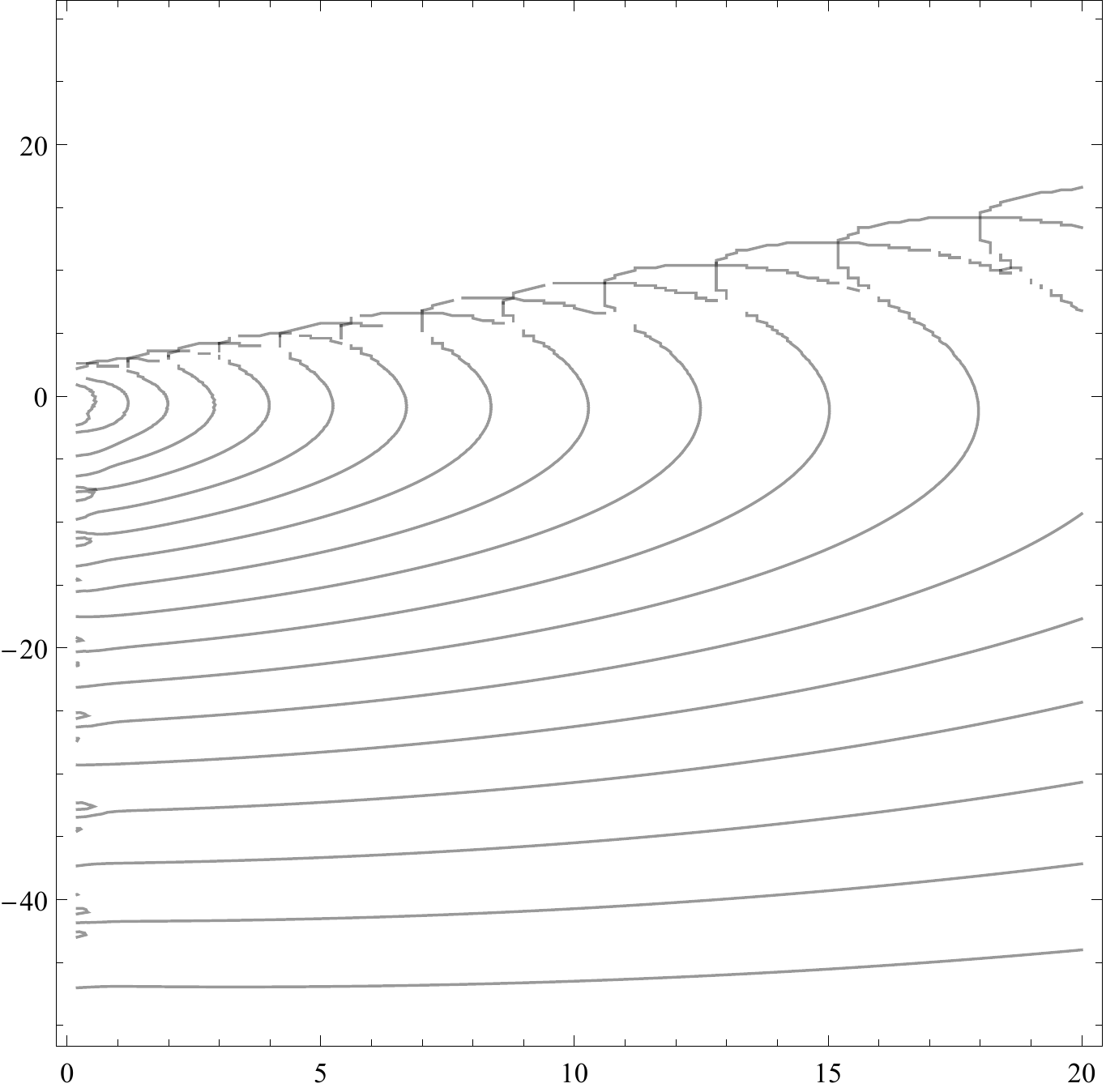}\\
\caption{$\xi=0.0,\ r_+=1.10$ and $\lambda=1.5$ (left), $\lambda=3.0$ (right)} \label{fig7}
\end{figure}

\subsection{QNMs for $r_+=1.10$ and various values of $\xi.$}

Fig.~\ref{fig4} contains the QNMs when $\lambda=0.5$ is small, $r_+=1.10$ and $\xi$ is set either to $0.0$ or to $5.0.$ The influence of the value of $\xi$ is apparent: the real parts of the QNMs move towards bigger positive values.

\begin{figure}[H]
\centering
\includegraphics[scale=0.4]{lambda05_xi0_rp110.pdf}
\includegraphics[scale=0.4]{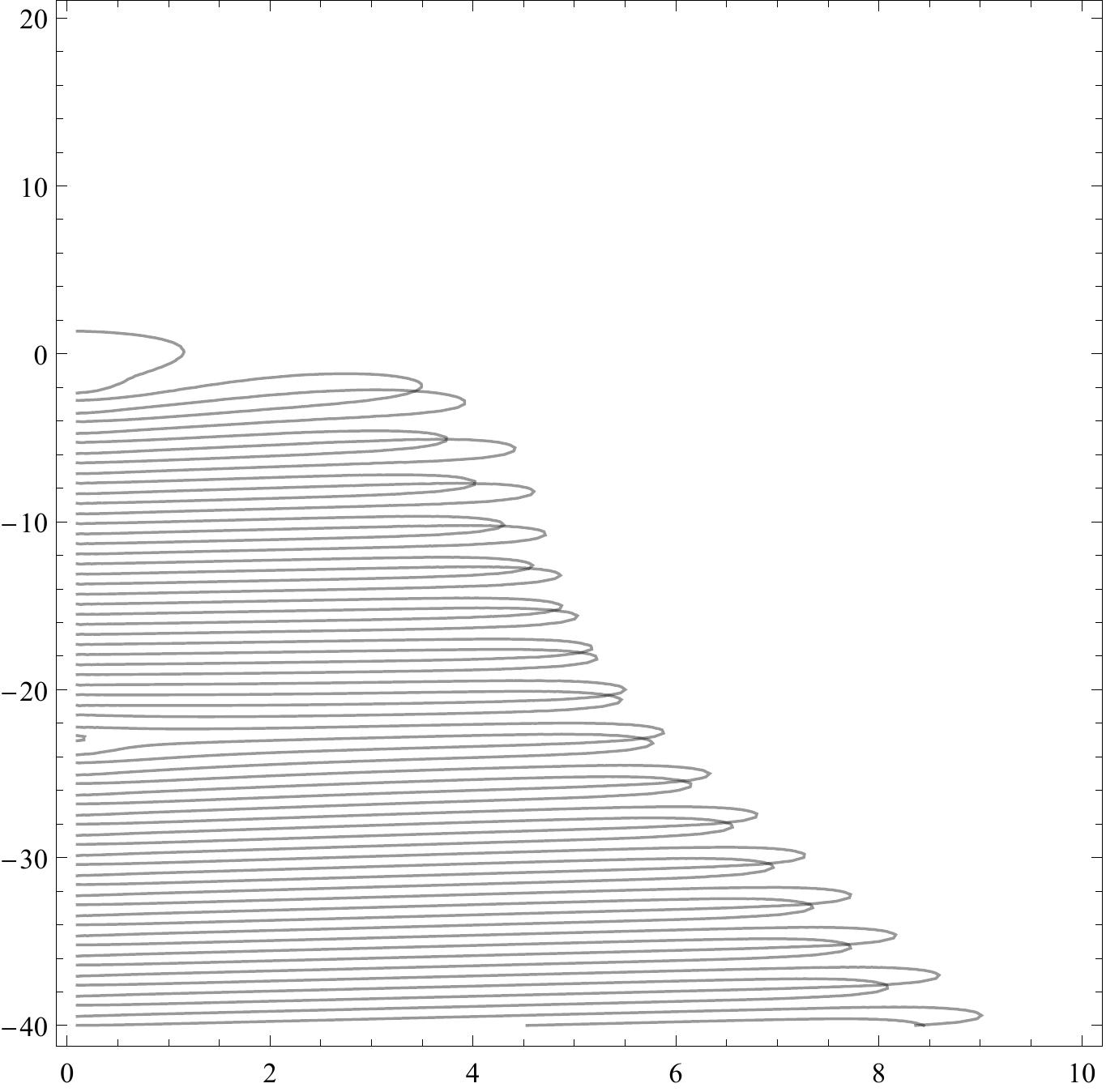}\\
\caption{$\lambda=0.50,\ r_+=1.10$ and $\xi=0.0$ (left), $\xi=5.0$ (right).} \label{fig4}
\end{figure}

On the other hand the influence of $\xi$ is somewhat different when $\lambda=1.5,$ that is when it takes a moderately large value. The situation for $\xi=0$ is depicted in Fig.~\ref{fig3}, left panel. In Fig.~\ref{fig6}, left panel, one may see the modifications brought about by the increasing values for $\xi:$ when $\xi=5.0,$ a modest value, the QNM's with negative $\omega_I$ are not modified very much; on the contrary the QNMs with positive $\omega_I$ are influenced. The nature of this change becomes clear for the value $\xi=30.0,$ shown in Fig.~\ref{fig6}, right panel: the QNMs with positive $\omega_I$ disappear completely, while the QNMs with negative $\omega_I$ move to less negative values. Once more, the real part of the QNMs moves to values of the order of $\xi.$ Thus the unstable system depicted in Fig.~\ref{fig3}, left panel is transformed through the situation in Fig.~\ref{fig6}, left panel, to the stable system shown in Fig.~\ref{fig6}, right panel. Thus increasing the $\xi$ value counterbalances the instability. In general it seems that the parameters $\lambda$ and $\xi$ act competitively. Looking at this behaviour another way, we find out that there is a critical value of $\xi$ for each value of $\lambda,$ such that below it the system is unstable.

\begin{figure}[H]
\centering
\includegraphics[scale=0.4]{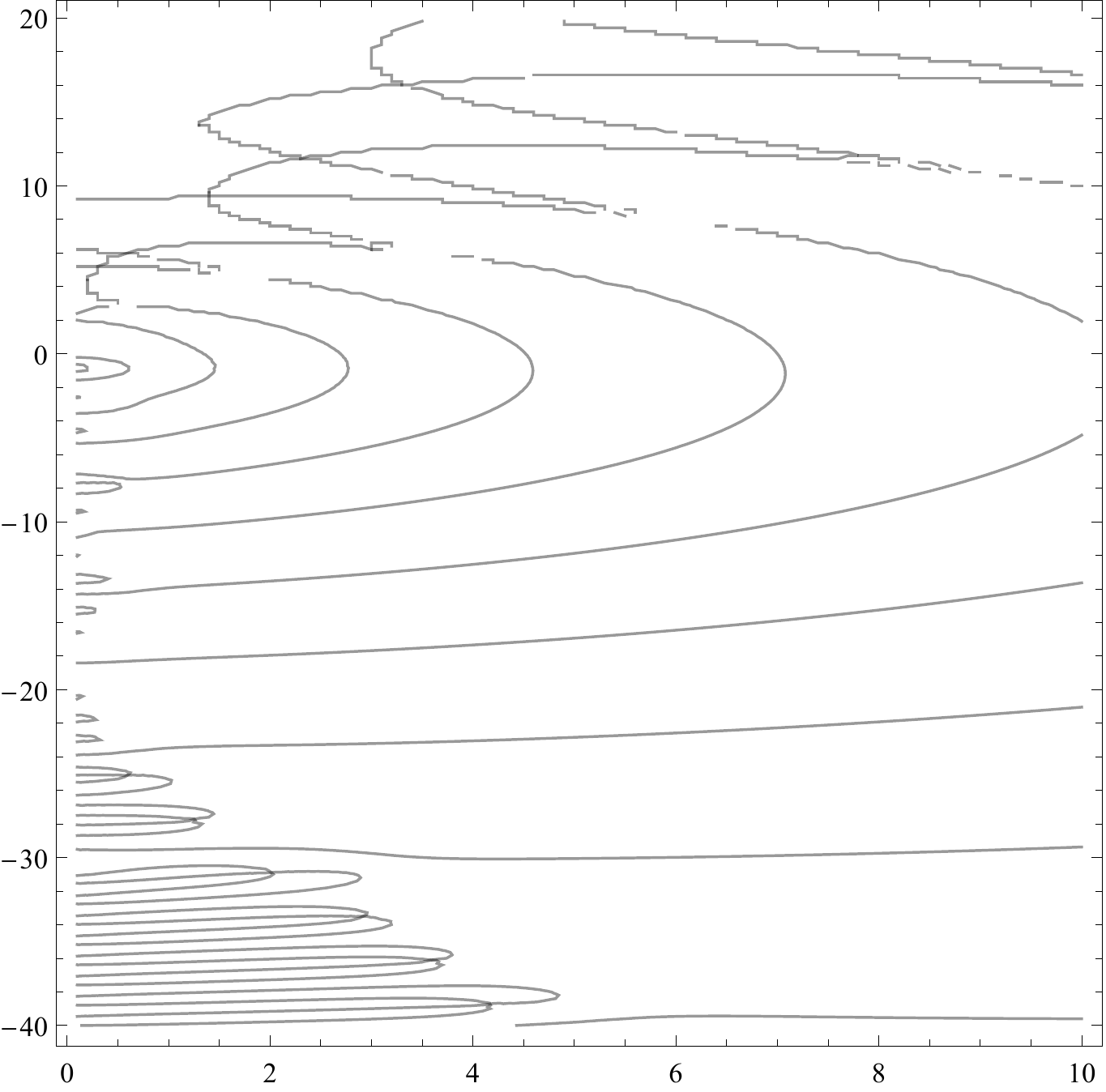}
\includegraphics[scale=0.4]{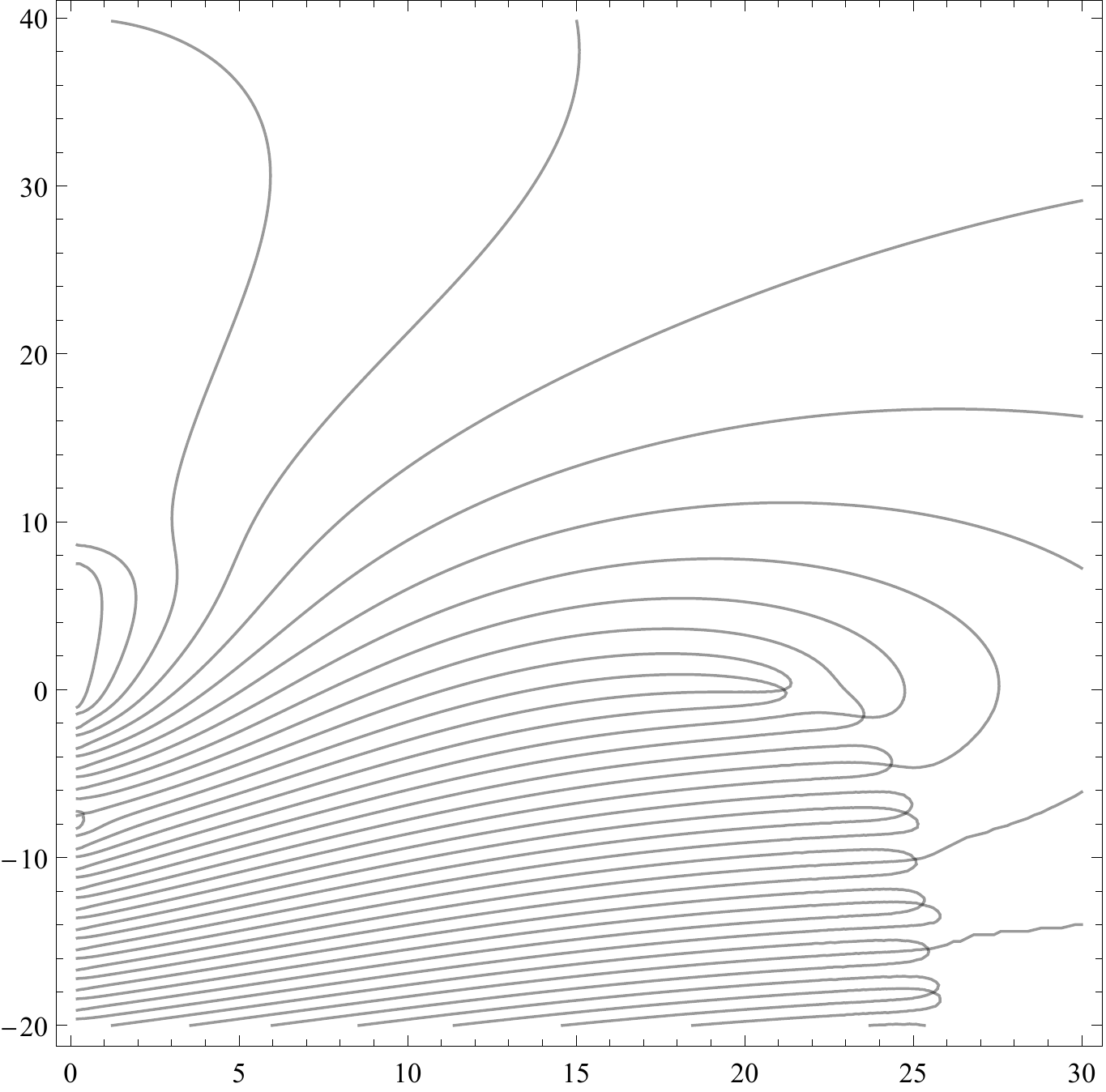}\\
\caption{$\lambda=1.50,\ r_+=1.10$ and $\xi=5.0$ (left), $\xi=30.0$ (right)} \label{fig6}
\end{figure}

\section{Conclusions}
\label{sec5}

 In this work we studied the stability of a topological black hole in the presence of the coupling of a scalar field to the GB term in  the ESTGB gravity theories and we investigated the possibility of its scalarization to an MTZ-like black hole. We considered a gravity theory in the presence of a cosmological constant and a massive scalar field  minimally coupled to gravity and also coupled to the GB term. The coupling of the scalar field to the GB term is denoted by the parameter $\lambda$. We perturbed the scalar field coupled to the GB term in the background of the  topological black hole living in a hyperbolic space-time expressed by the parameter $\xi.$

 We calculated both analytically and numerically the QNMs of scalar perturbations of topological-AdS black holes  with the purpose to study the behaviour of matter in this physical set up.  For fixed cosmological constant we have two competing effects. The first one is as $\lambda$ is increasing we expect the matter to interact more strongly with gravity while as $\xi$ is getting larger the effects of the variations of the wave functions are dominant. Our goal was to see what are the effects of the increase of the strength of the   parameters $\lambda$  and $\xi$ and their possible interplay on the stability of the topological black hole and if there are indications of a phase transition to a new scalarized black hole.

We first calculated analytically the QNMs of scalar perturbations. For small topological black holes (we had fixed the horizon radius to $r_{+}=1$) we found a critical value of $\lambda$ below which the topological black hole is stable under scalar perturbations. When the coupling constant $\lambda$  is getting  larger than its critical value, all of the QNMs develop a positive imaginary part signalling  an instability of the background black hole. Calculating the scalar modes of the perturbations we found that for large $\xi$  the variations of the wave functions influences most effectively the behaviour of the QNMs.

Calculating the QNMs of scalar perturbations numerically we get similar results for the instability of the background topological black hole. However, we get a better insight of the role of the hyperbolic geometry. For a fixed value of $\lambda$ above its critical value, the increase of the $\xi$ value counterbalances the instability which leads to a very interesting behaviour,  the parameters $\lambda$ and $\xi$ act competitively. Looking at a  different way, we found that there is a critical value of $\xi$ for each value of $\lambda,$ such that below this value the system is unstable.

To summarize our results we found that there are critical values of the parameter $\lambda$, which is the strength of the coupling of matter to the GB term, and the parameter $\xi$ which specifies the  geometry of the background metric,  which controls the instability of the topological black hole. Therefore we expect that the interplay of these parameters will lead to the scalarization of the topological black hole. To find the form of the scalarized topological black hole we have to allow the back-reaction of the scalar field to the background topological black hole. We leave this for future work.

\end{document}